\documentclass[fleqn,usenatbib]{mnras}

\usepackage{newtxtext,newtxmath}
\usepackage{pstricks,psfrag}
\usepackage{fancybox}
\usepackage[T1]{fontenc}

\DeclareRobustCommand{\VAN}[3]{#2}
\let\VANthebibliography\thebibliography
\def\thebibliography{\DeclareRobustCommand{\VAN}[3]{##3}\VANthebibliography}

\usepackage{graphicx}	

\newcommand{\dif}{{\rm d}}
\newcommand{\del}{\partial}
\newcommand{\veps}{\varepsilon}
\newcommand{\vphi}{\varphi}
\newcommand{\hN}{\hat{N}}
\newcommand{\hM}{\hat{M}}

\newcommand{\bb}{\mathbf{b}}
\newcommand{\bx}{\mathbf{x}}
\newcommand{\bp}{\mathbf{p}}

\newcommand{\by}{\mathbf{y}}
\newcommand{\bz}{\mathbf{z}}
\newcommand{\bF}{\mathbf{F}}

\title[A novel formulation for the evolution of relativistic rotating stars]{
A novel formulation for the evolution of relativistic rotating stars
}

\author[H. Okawa et al.]{
Hirotada Okawa,$^{1}$\thanks{E-mail: h.okawa@aoni.waseda.jp (WIAS)}
Kotaro Fujisawa,$^{2}$
Nobutoshi Yasutake,$^{3,4}$
Misa Ogata,$^{5}$
Yu Yamamoto,$^{5}$
and Shoichi Yamada$^{5}$
\\
$^{1}$Waseda Institute for Advanced Study(WIAS), 1-21-1 Nishi-Waseda,
Tokyo 169-0051, Japan\\
$^{2}$Department of Physics, Graduate School of
Science, University of Tokyo, Bunkyo-ku, Tokyo 113-0033, Japan\\
$^{3}$Physics Department, Chiba Institute of Technology, Chiba 275-0023,
Japan\\
$^{4}$Advanced Science Research Center, Japan Atomic Energy Agency,
Tokai, Ibaraki 319-1195, Japan\\
$^{5}$Research Institute for Science and Engineering, Waseda University,
Tokyo 169-8555, Japan
}

\date{Accepted XXX. Received YYY; in original form ZZZ}

\pubyear{2022}

\begin{document}
\label{firstpage}
\pagerange{\pageref{firstpage}--\pageref{lastpage}}
\maketitle

\begin{abstract}
We present a new formulation to construct 
numerically equilibrium configurations of rotating stars in general relativity.
Having in mind the application to their
quasi static evolutions,
we adopt a Lagrangian formulation of our own devising,
in which we solve force balance equations
to seek for the positions of fluid elements
assigned to the grid points,
instead of the ordinary Eulerian formulation.
Unlike previous works in the literature,
we do not employ the first integral
of the Euler equation,
which is not obtained by an analytic integration in general.
We assign a mass, specific angular momentum
and entropy to each fluid element in contrast to
the previous methods,
in which the spatial distribution of the angular velocity
or angular momentum is specified.
Those distributions are determined
after the positions of all fluid elements
~(or grid points) are derived in our formulation.
We solve the large system of algebraic nonlinear equations
that are obtained by discretizing the time-independent
Euler and Einstein equations
 in the finite-elements method by using
our new multi-dimensional root-finding scheme, named the W4 method.
To demonstrate the capability of our new formulation,
we construct some rotational configurations both
 barotropic and baroclinic.  
We also solve 
three evolutionary sequences that mimic the
cooling, mass-loss, and mass-accretion as simple toy models.
\end{abstract}




\section{Introduction}\label{sec:intro}
%
Neutron stars are promising compact objects
to obtain the information on
 the equation of state of dense matter
 and investigate physics of strong gravity.
The direct detections of gravitational waves
 from merging compact objects
 opened up a new window to access
 such information through the observation 
 (\citealp{PhysRevLett.116.061102,PhysRevLett.119.161101,Abbott_2021}).
The neutron star is born in core-collapse supernova
~(for a review see, \citealp{Janka2012,Burrows:2020qrp}).  
A successful explosion leaves a proto-neutron star~(PNS)
in its supernova remnant,
being decoupled from the expanding ejecta.
At its birth, the PNS is initially hot
and proton-rich (that is why it is called the proto-neutron star)
and has a radius a few times larger than the typical 
 size, $10 \mathrm{km}$, of neutron stars.
Neutrinos and gravitational waves
carry away energy from the PNS and cool it down
 to the neutron star on a diffusion timescale~(a few tens of seconds) much longer
than the dynamical timescale~(\citealp{Prakash:1996xs}).

It is quite numerically costly and impossible to follow the whole evolution
 of the PNS cooling
 by a dynamical simulation in multi-dimensions.
Even recent core-collapse supernova simulations
covered only the first
$10$ \textrm{seconds} at most
~(\citealp{Muller:2018utr,Nakamura:2019snn,Burrows:2019zce,Nagakura:2017mnp}).
Thanks to the large separation of the two timescales,
this long, quasi-static evolution of the PNS
cooling may be approximated in a sequence of
equilibrium configurations with gradually changing thermal
and lepton contents.
In fact, this was the common strategy in the past
~(\citealp{Burrows:1986me,Keil:1996ab,Pons:1998mm}).
Note that this was made possible by the fact
that the Lagrangian formulation is easily employed
in spherical symmetry.

It is not a trivial task, however,
to adopt the same strategy in multi-dimensions
for rotating stars.
One of the big issues to be resolved is
to devise a Lagrangian formulation that enables us
to construct the equilibrium configurations of rotating stars.
To the best of our knowledge, there has been no such method
so far except for our own previous attempt in
Newtonian gravity~(\citealp{Yasutake:2015,Yasutake:2016}),
in which we employ a triangular Lagrangian grid.
It turns out, unfortunately, this scheme was neither robust
nor very efficient.
We hence need to 
build a new numerical scheme that
is suitable for the study of the secular evolution
of relativistic rotating stars such as PNSs
to construct stationary rotating stars
and solve two major issues:
(i) to conceive a Lagrangian formulation in general relativity
and (ii) its robust and efficient implementation\footnote{
A relativistic smoothed-particle-hydrodynamics formulation has been recently considered
as dynamical problems in~\citet{Rosswog:2020kwm,Diener:2022hui},
whereas the Lagrange formulation as boundary value problems is not investigated in general relativity so far.}.
The latter is rather technical but turns out to be
crucial as we explain later in detail.


The equilibrium configurations of
relativistic rotating stars
have been extensively investigated
in the Eulerian formulation
 so far~(see \citealp{Paschalidis:2016vmz} for review):
Rotating solutions in general relativity were discussed by \cite{Hartle:1967} and
\cite{Butterworth:1976},
differentially rotating stars were  
successfully constructed by~\citet{Komatsu1989b,Cook:1992}
\footnote{This scheme was implemented in the RNS code by~\cite{Stergioulas1995} which is now put in the public domain.};
higher accuracy was attained by a 
spectral solver in~\cite{Bonazzola:1993} and
~\cite{Ansorg:2002}.
More recent progresses now allow triaxial stars and magnetized stars
~(\citealp{Uryu:2011ky,Zhou:2017xhf,Uryu:2019ckz}).
Note, however, that all these works made two assumptions \textit{a priori}:
(i) the barotropic condition, i.e., the pressure depends only on the density and
(ii) the rotation law that assumes a functional relation between specific angular momentum and angular velocity, i.e., $F=F(\Omega)$~(\citealp{Komatsu1989b}) or $\Omega=\Omega (F)$~(\citealp{Uryu:2017obi}).
Under these assumptions, one can analytically
integrate the Euler equation and the main task is
to solve the Einstein equations.
Non-barotropic stars in general relativity
have been also built, though.
In general relativity, for example,
\cite{Camelio:2019rsz} proposed 
a restrictive formulation by introducing
$p-\Omega$ coordinates
whereas there has been more
works in the Newtonian gravity
~(\citet{Uryu:1994,Uryu:1995,Roxburgh2006,Espinosa-Lara2007,Espinosa-Lara2013,Yasutake:2015,Yasutake:2016,Fujisawa:2015}).

Based on the above Eulerian formulation of rotating stars,
many authors have attempted to describe rotating PNSs
~(\citet{Goussard:1996dp,Goussard:1997bn,Sumiyoshi1999,Strobel:1999vn,Villain:2003ey,Camelio:2016fan}).
Thanks to sophistication of numerical relativity
with the rapid increase in computational power,
on the other hand,
some groups have started to use dynamical simulations~(\citet{Camelio:2019rsz,Zhou:2021upu,Fujibayashi:2021wvv})
although they are limited to rather short periods.
The dynamical simulations are put aside
and the Eulerian formulation faces
a difficulty when
applied to the evolution of rotating stars:
the angular momentum distribution \textit{in space}
is not known a priori~(see assumption (ii) above)
as a function of time even if there is no angular momentum
transfer in the star.
This is because in such a situation
the specific angular momentum is conserved
for the individual fluid element but the angular momentum distribution
in space changes in time.
This problem is automatically solved 
if one employs a Lagrangian formulation
although it is highly nontrivial.
That is our core idea in this paper.

The organization of the paper is as follows.
In Sec.~\ref{sec:method}, we 
present this Lagrangian formulation of our devising.
We then explain rather in detail how this formulation is implemented as a numerical scheme, since this part is actually crucial.
We describe the models constructed in this paper
to demonstrate the capability of our new method
 in Sec.~\ref{sec:model}
and show the results in Sec.~\ref{sec:result}.
Finally, we summarize our findings and give future prospects in Sec.~\ref{sec:conclusion}.
Throughout this paper, geometrized units, i.e., $c=G=1$ are used unless otherwise noted.

\section{Method}\label{sec:method}
%
%
We start with a brief review of the ordinary numerical
construction of axisymmetric rotating stars in
general relativitic gravity in e.g.~\citet{Komatsu:1989,Cook:1992,Cook:1993qj}.
Under axisymmetry and stationarity without circulation
flows\footnote{The metric ansatz for the case 
with circulation flows is
discussed in \cite{Birkl:2010hc}.  In this case, 
there are eight metric functions instead of four.}
the line element is given in general as
\begin{equation}
 \dif s^2 = -N^{2} \dif t^2
  +A^{-2}\left(\dif r^2 +r^2\dif \theta^2\right)
  +B^{-2}r^2\sin^2\theta
  \left(\dif\vphi -\omega\dif t\right)^2,
  \nonumber\\
\end{equation}
where $N(r,\theta), A(r,\theta), B(r,\theta), $ and $\omega(r,\theta)$
are functions of the radius and the zenith angle.
Note that the metric components are not written 
in the exponential form
~(\cite{Komatsu1989b}) but in the $3+1$ style in~\cite{Bonazzola:1993}.
The matter is assumed to be a perfect fluid
and its energy momentum tensor is given as
\begin{eqnarray}
 T_{\mu\nu} = \left(\veps+P\right)u_{\mu}u_{\nu} +Pg_{\mu\nu},
\end{eqnarray}
where $u^{\mu}$ is the 4-velocity and $\veps \equiv \rho +\rho_{th}$
is the energy composed of the rest mass density~$\rho$ and the internal energy density~$\rho_{th}$.
The four-velocity is given as
\begin{eqnarray}
 u^{\mu}=\left(\frac{1}{N\sqrt{1-v^2}},
	  0,
	  0,
	  \frac{\Omega}{N\sqrt{1-v^2}}\right)^{T},
\end{eqnarray}
where $v=\left(\Omega-\omega\right)N^{-1}B^{-1}r\sin\theta$
and $u^{\mu}u_{\mu}=-1$.
The angular velocity observed by the Zero
Angular Momentum Observer~(ZAMO)
is denoted by $\Omega(r,\theta)$.

Four metric functions~$N(r,\theta), A(r,\theta), B(r,\theta)$
and $\omega(r,\theta)$ are determined
by solving the Einstein equation~$\displaystyle G_{\mu\nu}=8\pi T_{\mu\nu}$.
The energy-momentum conservation equation or the relativistic Euler equation is given by
$\nabla_{\nu}T_{\mu}^{\ \nu}=0$.
The explicit differential forms of all these equations are presented in App.~\ref{sec:diffeqs}.
If the matter is barotropic, i.e., its pressure depends only on the density~$P=P(\rho)$, the equation of state~(EOS) closes the system equation.
If the matter is baroclinic, i.e., the pressure depends
on another thermodynamic quantity, say the specific entropy~$s$ as $P=P(\rho,s)$\footnote{More generally, the pressure may depend on yet another thermodynamic quantity such as
the electron fraction: $P=P(\rho,s,Y_e)$.},
then the equation is augmented with the equation for energy.
One often needs to solve the radiation transport equation as well
~(\cite{Pons:1998mm}).

There is a successful strategy
 common to the methods proposed in the literatures to numerically construct general relativistic rotating stars~(see ~\cite{Paschalidis:2016vmz} for a recent review):
(i) the Einstein equation is first solved for a given matter
configuration; fixing the metric so obtained, we solve
 the Euler equation for density; replacing the matter configuration
employed in the first step with the density distribution obtained in the second step,
we iterate these two steps
until the convergence is achieved
as the self-consistent field method for Newtonian rotating stars in ~\cite{Hachisu:1986};
(ii) the Euler equation is analytically integrated in advance,
which is actually possible for
the barotropic case: $\veps=\veps(P)$ or $P=P(\veps)$
with $F(\Omega)= u^{t}u_{\vphi}$ where $F(\Omega)$ is
an arbitrary function.
In fact, the Euler equation under those assumptions:
\begin{eqnarray}
 \frac{1}{\veps+P}\dif P -\dif \ln u^{t} +u^{t}u_{\vphi}\dif\Omega = 0,\label{eq:Euler}
\end{eqnarray}
leads to the first integral as follows:
\begin{eqnarray}
 \mathcal{H}(P) -\ln u^{t} +\mathcal{F}(\Omega) = \mathcal{C},
\end{eqnarray}
where $\mathcal{H}$ and $\mathcal{F}$ are given as
\begin{eqnarray}
 \mathcal{H}(P) \equiv \int^{P} \frac{\dif P'}{\veps(P')+P'},\quad
  \mathcal{F}(\Omega) \equiv \int^{\Omega}F(\Omega')\dif\Omega',
\end{eqnarray}
and $\mathcal{C}$ is an integration constant 
and may be determined at the pole.
Although the methods based on this strategy have been
successful in the numerical construction of a rotational
configuration for a given functional form of angular velocity~$F(\Omega)$, they are not suited for the study of
secular evolutions of rotating stars.
There are two problems in fact:
firstly, these methods are in principle applicable
only to the barotropic case; secondly,
the angular velocity distribution should be given in advance,
which is normally impossible.
The latter point will be understood
if one considers the secular thermal evolution
with the specific angular momentum
being conserved in each fluid element.
In fact, the spatial distribution of the angular velocity
changes in time in this case even though
there is no angular momentum transfer.
This last point motivated us to develop a Lagrangian method
in this paper, with which such a change of the angular velocity
distribution can be automatically solved.
Note also that it is not necessary to use the first integral
in our formulation~(see
\ref{eq:Euler_r} and \ref{eq:Euler_th}).


%
\subsection{Lagrangian method}
In the Eulerian formulation, one first introduces
the coordinates
rather arbitrarily according to the convenience
in their applications; once chosen,
they are unchanged; we then solve the
density(or pressure) 
and the metric functions of the coordinates so that
they should satisfy the Euler equation
and the Einstein equation, respectively.
As mentioned earlier,
the angular velocity profile, or equivalently the angular
momentum distribution should be given as input
normally.
It is also mentioned that 
the mass and angular momentum 
that characterizes a rotational star are normally obtained
only after the solution is obtained.

In sharp contrast, in the Lagrangian formulation we solve the Euler
equation to obrain the coordinates,
which are attached to fluid elements.
More specifically, assigning the mass, specific angular momentum
and specific entropy to each fluid element\footnote{The specific angular momentum and specific entropy hence regarded as functions of fluid elements. We may also give the electron fraction~$Y_e$.},
we will look for the positions of fluid elements that satisfy
the Euler equation.  In so doing,
the density is a functional of the coordinate configurations.
The pressure at each fluid element is derived with the equation
of state once the density is known there as the specific entropy
is also known \textit{a priori}.
The spatial distribution of the angular velocity,
or equivalently the angular momentum, is obtained
once the coordinate configuration is determined.
It should be apparent that the mass and angular momentum
are automatically conserved in this formulation.
In the following we will explain more in detail
how these procedures are realized and implemented
in the numerical construction of rotating stars.

\subsubsection{Density in a Finite Element}
\begin{figure}
 \psfrag{x}{$x$}
 \psfrag{z}{$z$}
 \psfrag{r11}{\small \hspace{-5mm}$(r_{11},\theta_{11})$}
 \psfrag{r12}{\small \hspace{0mm}$(r_{12},\theta_{12})$}
 \psfrag{r13}{\small \hspace{0mm}$(r_{13},\theta_{13})$}
 \psfrag{r14}{\small \vspace{-5mm}$(r_{14},\theta_{14})$}
 \psfrag{r21}{\small \hspace{0mm}$(r_{21},\theta_{21})$}
 \psfrag{r22}{\small \vspace{5mm}$(r_{22},\theta_{22})$}
 \psfrag{r23}{\small \hspace{0mm}$(r_{23},\theta_{23})$}
 \psfrag{r24}{\small \hspace{-3mm}$(r_{24},\theta_{24})$}
 \psfrag{rho11}{\small \hspace{0mm}$\boxed{\bar\rho_{11}}$}
 \psfrag{rho12}{\small \hspace{0mm}$\boxed{\bar\rho_{12}}$}
 \psfrag{rho13}{\small \hspace{0mm}$\boxed{\bar\rho_{13}}$}
 \psfrag{rho21}{\small \hspace{0mm}$\boxed{\bar\rho_{21}}$}
 \psfrag{rho22}{\small \hspace{0mm}$\boxed{\bar\rho_{22}}$}
 \psfrag{rho23}{\small \hspace{0mm}$\boxed{\bar\rho_{23}}$}
 \includegraphics[width=7.5cm]{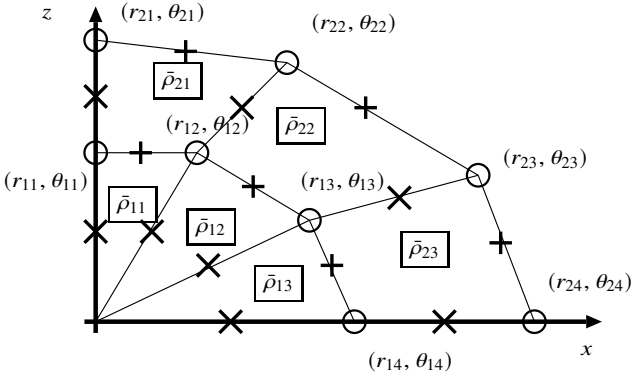}
 \caption{Schematic picture of finite elements.
 The circles denote the Lagrange nodes
 and the evaluation point for the Euler equation.
 Other physical variables are defined at the crosses
 and the Einstein equation is solved there.
 }
 \label{fig:def_FE}
\end{figure}
%
In our brand new Lagrangian formulation of 
axisymmetric stars in permanent rotation,
the starting point is to express the density
in terms of the coordinates configuration.
For its numerical realization on the finite-elements~(FEs),
we approximate each fluid element in the meridian section by
a quadrilateral FE and assign
mass~$\Delta m$, specific angular momentum~$\Delta j$, specific entropy~$\Delta s$,
and electron fraction~$\Delta Y_e$, etc.
The configuration of FE is specified by the coordinates~$(x_{jk}, y_{jk})$, or specifically $(r_{jk},\theta_{jk})$ in axisymmetric stars, of its four corners
as shown in Fig.~\ref{fig:def_FE}
and its area may be calculated conveniently with the isoparametric
formulation as follows~(see, e.g., \cite{bathe2006finite}).
In this formulation, we 
introduce the natural coordinates~$(\alpha,\beta)$,
 $-1\ge \alpha, \beta \ge 1$, which specify an arbitrary point
in the FE as
\begin{eqnarray}
 x(\alpha,\beta) &=& \sum_{j=1}^{2}\sum_{k=1}^{2}
  \hN_{j}(\alpha)\hN_{k}(\beta)x_{jk},\nonumber\\
 y(\alpha,\beta) &=& \sum_{j=1}^{2}\sum_{k=1}^{2}
  \hN_{j}(\alpha)\hN_{k}(\beta)y_{jk},\label{eq:map_FE1}
\end{eqnarray}
where $\hN_{j} (j=1,2)$ are called the shape function;
$x_{jk}$ and $y_{jk}$ are the original coordinates
at the four corners.
Equation~\eqref{eq:map_FE1} is actually an interpolation
formula and the linear shape functions are given as
\begin{eqnarray}
  \hN_{1}(\alpha) = \frac{1-\alpha}{2},\quad
  \hN_{2}(\alpha) = \frac{1+\alpha}{2}.
\end{eqnarray}
In App.~\ref{sec:isoparametric},
we explain the basic idea underlying this formulation.
The coordinate transformation from $(x,y)$ to $(\alpha,\beta)$
is characterized by
the Jacobian matrix
\begin{eqnarray}
 J =
   \begin{pmatrix}
    \displaystyle \frac{\del x}{\del \alpha} &
    \displaystyle \frac{\del y}{\del \alpha}\\
    & \\
    \displaystyle \frac{\del x}{\del \beta} &
    \displaystyle \frac{\del y}{\del \beta}
   \end{pmatrix}.
\end{eqnarray}
For instance, the elements of which are given as
\begin{eqnarray}
 \frac{\del x}{\del \alpha} =  \sum_{j=1}^{2}\sum_{k=1}^{2}
  \frac{\dif \hN_{j}}{\dif \alpha}(\alpha)\hN_{k}(\beta)x_{jk}.
\end{eqnarray}
Adopting $x_{jk}=r_{jk}^3/3$ and $y_{jk}=\cos\theta_{jk}$
in the current case,
we define the volume of the FE in the flat space as
\begin{eqnarray}
 \Delta V = 2\pi |\det J|.
\end{eqnarray}
By accounting for the curvature of the space,
 the baryonic density for the FE is given by
\begin{eqnarray}
 \bar{\rho} = \frac{\Delta m}{\sqrt{-g}\Delta V},
  \label{eq:density_def}
\end{eqnarray}
where the $g$ is the determinant of the spacetime metric.
Since we evaluate the Einstein equation at each
cell boundary as marked with crosses in Fig.~\ref{fig:def_FE},
we need the densities at these boundaries.
In this paper, they are given as a simple arithmetic mean
of the densities for the adjascent FEs as
$\rho_{jk} \equiv \left(\bar{\rho}_{jk} + \bar{\rho}_{jk-1}\right)/2$.
Note again that the density itself is the variable to be solved
in the Eulerian formulation, whereas
the coordinates are to be solved
in the Lagrangian formulation.

\subsubsection{Isoparametric interpolation and differentiation}
In order to solve the Einstein and Euler equations in their differential forms, Eqs.~\eqref{eq:Euler_r}-\eqref{eq:Etph}, in the Lagrangian formulation,
we need to evaluate not only the values of variables but also their derivatives at an arbitrary position.
It is not a difficult task in the FE description.
Since we need to evaluate second-order derivatives in the
Einstein equation, we employ the second-order interpolation,
in which we use quadratic shape functions
and not the values of four but nine nearby points.
Then the coordinates are expressed in terms of the natural coordinates~$\alpha$ and $\beta$ as
\begin{eqnarray}
 x(\alpha,\beta) &=& \sum_{j=1}^{3}\sum_{k=1}^{3}
  \hM_{j}(\alpha)\hM_{k}(\beta)x_{jk},\label{eq:xiso}\\
 y(\alpha,\beta) &=& \sum_{j=1}^{3}\sum_{k=1}^{3}
  \hM_{j}(\alpha)\hM_{k}(\beta)y_{jk},\label{eq:yiso}\
\end{eqnarray}
where the shape functions are given by
\begin{eqnarray}
  \hM_{1}(\alpha) &=& -\frac{\alpha}{2}(1-\alpha),\nonumber\\
  \hM_{2}(\alpha) &=& (1+\alpha)(1-\alpha),\nonumber\\
  \hM_{3}(\alpha) &=& \frac{\alpha}{2}(1+\alpha).
\end{eqnarray}
We expand any function~$\phi(x(\alpha,\beta),y(\alpha,\beta))$ 
with the same shape functions as
\begin{eqnarray}
 \phi (\alpha,\beta) &=& \sum_{j=1}^{3}\sum_{k=1}^{3}
  \hM_{j}(\alpha)\hM_{k}(\beta)\phi_{jk}.
\label{eq:fiso}
\end{eqnarray}
It is straightforward to evaluate the derivatives of such a function
with the derivatives of the shape functions:
\begin{eqnarray}
 \frac{\dif\hM_{1}}{\dif \alpha}(\alpha)
  &=& -\frac{1}{2} +\alpha,\nonumber\\
 \frac{\dif\hM_{2}}{\dif \alpha}(\alpha)
  &=& -2\alpha,\nonumber\\
 \frac{\dif\hM_{3}}{\dif \alpha}(\alpha)
  &=& \frac{1}{2} +\alpha.
\end{eqnarray}
For instance, the first derivative with respect to $x$ can be obtained at any point as
\begin{eqnarray}
 \frac{\del \phi}{\del x}
 &=& \sum_{j,k=1}^{3}
 \left\{ \frac{\del\alpha}{\del x}\frac{\dif \hM_{j}}{\dif \alpha}\hM_{k}
  +\frac{\del\beta}{\del x}\hM_{j}\frac{\dif \hM_{k}}{\dif \beta}
 \right\}\phi_{jk},
\end{eqnarray}
where the coefficients~$\displaystyle\frac{\del\alpha}{\del x}$
and~$\displaystyle\frac{\del\beta}{\del x}$ are 
computed straightforwardly from Eq.~\eqref{eq:xiso}
as the coefficients of the inverse of the Jacobian matrix:
\begin{eqnarray}
 J^{-1} =
   \begin{pmatrix}
    \displaystyle \frac{\del \alpha}{\del x} &
    \displaystyle \frac{\del \beta}{\del x}\\
    & \\
    \displaystyle \frac{\del \beta}{\del y} &
    \displaystyle \frac{\del \alpha}{\del y}
   \end{pmatrix}
   =
   \frac{1}{\det J}
   \begin{pmatrix}
    \displaystyle \frac{\del y}{\del \beta} &
    \displaystyle -\frac{\del y}{\del \alpha}\\
    & \\
    \displaystyle -\frac{\del x}{\del \beta} &
    \displaystyle \frac{\del x}{\del \alpha}
   \end{pmatrix}.
\end{eqnarray}

\subsubsection{Angular momentum}
There are different definitions of specific angular momentum
employed in the literature.
For example, in the stationary and axisymmetric spacetime,
the following specific angular momentum~$\ell$ is introduced:
\begin{eqnarray}
 \label{eq:lmom}
 \ell = -\frac{u_{\vphi}}{u_t}
 = \frac{\left(\Omega-\omega\right) r^2\sin^2\theta}
 {N^2B^2+\omega\left(\Omega-\omega\right)r^2\sin^2\theta},
\end{eqnarray}
which is conserved along a stream line~(\cite{Birkl:2010hc}).
EoM, which can be written as
\begin{eqnarray}
 \nabla^{\nu} T_{\mu\nu} &=& 
  u^{\nu}\nabla_{\nu}\left[\left(\veps+P\right)u_{\mu}\right]
  +\left(\veps+P\right)u_{\mu}\nabla_{\nu}u^{\nu} +\nabla_{\mu}P
  \nonumber\\
 &=&
  u^{\nu}\nabla_{\nu}\left[\frac{\left(\veps+P\right)}{\rho}u_{\mu}\rho\right]
  -\frac{\left(\veps+P\right)}{\rho}u_{\mu}u^{\nu}\nabla_{\nu}\rho +\nabla_{\mu}P\nonumber\\
&=& \rho u^{\nu}\nabla_{\nu}\left[\frac{\left(\veps+P\right)}{\rho}u_{\mu}\right]
  +\nabla_{\mu}P = 0,
\end{eqnarray}
where the continuity equation is substituted, i.e.
~$\rho\nabla_{\nu}u^{\nu} = -u^{\nu}\nabla_{\nu}\rho$,
one finds for the spacetime with the asymptotically
timelike and axial Killing vectors
there are actually two quantities conserved along each stream line:
\begin{eqnarray}
 j_{t} = \frac{\veps+P}{\rho} u_{t}\quad \mathrm{and}\quad
 j_{\vphi} = \frac{\veps+P}{\rho} u_{\vphi}.
\end{eqnarray}
The specific angular momentum defined in Eq.~\eqref{eq:lmom}
is nothing but 
the ratio of these two:~$\ell=-j_{\vphi}/j_{t}$ 
Since we are interested in the formulation that
can be applied to the evolution of rotating stars,
the existence of the timeline Killing vector cannot be assumed.
It should be noted that $j_{\vphi}$ is still conserved
along the stream line even in this case
as long as the spacetime is axisymmetric.
We will hence employ this specific angular momentum
and assume that it is conserved for each fluid element
unless some mechanism to exchange angular momenta
between fluid elements is in operation.
Note, however, that
the specific angular momentum~$\ell$ is still very convenient
from numerical point of view.
In fact, it is simple
to convert $\ell$
to the angular velocity~$\Omega$
\begin{eqnarray}
 F\left(\Omega\right) = u^tu_{\vphi} = \frac{\ell}{1-\Omega\ell},
  \label{eq:Fomegaell}
\end{eqnarray}
where $u_{\mu}u^{\mu}=-1$ and $u^{\vphi}=\Omega u^{t}$.


%
\subsection{Diagnostics}
The following
 global quantities are useful for characterizing
rotational equilibria.
Following ~(\citet{Cook:1992,Nozawa1998,Paschalidis:2016vmz}), we define 
the baryon mass, proper mass and gravitational mass, respectively, as
\begin{eqnarray}
 M_b &=& 2\pi\int \frac{\rho }{A^2B\sqrt{1-v^2}}
  r^2\sin\theta\dif r\dif\theta,\\
 M_p &=& 2\pi\int \frac{\epsilon}{A^2B\sqrt{1-v^2}}
  r^2\sin\theta\dif r\dif\theta,\\
 M &=& 2\pi\int \frac{1}{A^2B^2}
  \left[ NB\left\{ \frac{\left(\epsilon  +P\right)\left(1+v^2\right)}{\left(1-v^2\right)}
			+2P\right\}\right.\nonumber\\
&&\left. \quad\quad\quad\quad\quad  +2rv\omega\sin\theta \frac{\left(\epsilon +P\right)}{1-v^2}
  \right] r^2\sin\theta\dif r\dif\theta.
\end{eqnarray}
On the other hand, the quantities employed to measure how fast the rotation is are
the total angular momentum, rotational energy
and gravitational energy, respectively, as
\begin{eqnarray}
 J &=& 2\pi\int \frac{\left(\epsilon +P\right)v}{A^2B^2\left(1-v^2\right)}
  r^3\sin^2\theta\dif r\dif\theta,\\
 T &=& 2\pi\int \frac{\left(\epsilon +P\right)v\Omega}{A^2B^2\left(1-v^2\right)}
  r^3\sin^2\theta\dif r\dif\theta,\\
 W &=& M_p +T -M.
\end{eqnarray}
\subsubsection{Isoparametric integration}
The integrations above are evaluated in each FE
and summed as follows:
\begin{eqnarray}
 S &\equiv& \int\!\!\!\!\int\!\!\! \phi(x,y)\dif x\dif y
  = \int\!\!\!\!\int\!\!\! \phi\left(x(\alpha,\beta),y(\alpha,\beta)\right)
  |\det J|\dif\alpha\dif\beta\nonumber\\
 &=& \int\!\!\!\!\int\!\!\! \phi
  \left|\frac{\del x}{\del\alpha}\frac{\del y}{\del\beta}
  -\frac{\del x}{\del\beta}\frac{\del y}{\del\alpha}\right|
  \dif\alpha\dif\beta\nonumber\\
 &=& \sum_{i,j,k,l,m,n}^{2}\!\!\!\!\!\!
  \frac{\phi_{ij} x_{kl} y_{mn}}{4}\left|(-1)^{k+n}\mathcal{N}_{im}\mathcal{N}_{jl}
	      -(-1)^{m+l}\mathcal{N}_{ik}\mathcal{N}_{jn}
	     \right|,\nonumber\\
\end{eqnarray}
where the shape functions are analytically integrated
 over the ranges $-1\leq \alpha \leq 1$
 and $-1\leq \beta \leq 1$ in advance as
\begin{eqnarray}
 \mathcal{N}_{ij} = \frac{1}{6}
 \begin{pmatrix}
  1 & 2\\
  2 & 1
 \end{pmatrix}.
\end{eqnarray}

\subsection{Strategy to solve the system}
To solve the whole system, we adopt the traditional iterative scheme~(\citealp{Hachisu:1986,Komatsu:1989}).
Specifically, it consists of two parts:
(i) solving the Einstein equation, Eqs.~\eqref{eq:Ett}-\eqref{eq:Etph}, with the matter quantities unchanged
and (ii) solving the Euler equation, Eqs.~\eqref{eq:Euler_r}
 and~\eqref{eq:Euler_th}, with the metric functions fixed.
Both equations in their differential forms are
discretized on the FE grid and the resultant nonlinear
algebraic equations are solved alternately
until the changes in the solutions for
both equations become smaller than certain values.
As may be understood from the fact that many efforts have been made to
choose a nice set of equations even in the Eulerian formulation,
these nonlinear equations are difficult to solve and,
it turns out,
the original Newton-Raphson method does not work for our equations.

In order to tackle this problem, we deploy two new schemes
of our own devising:
(i) for the Einstein equation,
the W4IX method is applied;
as described in App.~\ref{app:W4IX},
it is an extension of the original W4 method
and requires only $\mathcal{O}(N^2)$ calculations to obtain a solution, in contrast to $\mathcal{O}(N^3)$
operations in 
 the original W4 method with either the UL or LH decomposition
~(see \citet{Okawa:2018smx,Fujisawa:2018dnh} for details);
(ii) for the Euler equation on the other hand, 
we find that a particular iteration scheme dubbed
the slice shooting is very powerful to obtain convergence;
as demonstrated 
in App.~\ref{app:sliceshooting},
the Jacobian matrix for the nonlinear force-balance equations
in the $\theta$-direction is especially ill-conditioned
as the size of the matrix becomes bigger;
we find it better to solve 
a single radial slice at a time
with all the other slices fixed;
we repeat this for all the slices, starting from the axis,
proceeding to the equator and going back to the axis;
this cycle is repeated until the convergence is obtained
globally.

These two schemes turn out to be very successful.
They are robust and also efficient and actually the key
ingredients of our new formulation.

\section{Models}\label{sec:model}
In this section, we describe some models we employ in this paper
to demonstrate the capability of our new method.
As explained already, in this formulation
we first assign 
the mass and specific angular momentum
to the grid points and
find their configuration
that satisfies the Euler equation, or
force-balance equations,
as well as the spacetime metric that is consistent with the matter
distribution by solving the Einstein equation.
The spatial profiles of the density
and angular velocity, respectively,
are obtained from Eqs.~\eqref{eq:density_def} and~\eqref{eq:lmom},
respectively,
 after the equilibrium configurations of
the matter and spacetime are derived.
Having an application to PNSs in mind,
we also allocate the electron fraction to the Lagrange grid points.

The ultimate goal of this project is to 
study secular evolutions of relativistic rotating stars.
The models are hence divided into two groups:
\begin{itemize}
 \item Stationary rotating stars constructed either
       with a barotropic EOS or with a baroclinic EOS
       and their accuracies investigated closely with a couple of
       diagnostic quantities; a comparison made with an Eulerian code,
 \item mock evolutionary sequences of rotating stars 
       considered for some scenarios:
       cooling, mass-loss, and mass-accretion;
       they are meant for demonstrations.
\end{itemize}
Each model is described more in detail below.

\subsection{Stationary models}
Here we adopt the so-called $j-$const law~(\citealp{Komatsu1989b}).
This is actually the easiest case for our formulation,
in which the specific angular momentum is assigned
to the grid points and fixed.
Then the angular velocity profile is derived
after the equilibrium configuration is obtained.
We begin with a barotropic EOS. In this case,
as mentioned in introduction,
we analytically obtain a first integral of the Euler equation,
which gives for the present case the relation
 that the angular velocity should satisfy as
\begin{eqnarray}
 F(\Omega) &\equiv& u^{t}u_{\vphi}\nonumber\\
 &=& A_c^2(\Omega_c-\Omega) =
  \frac{(\Omega-\omega)r^2\sin^2\theta }
  {N^2B^2-(\Omega-\omega)^2r^2\sin^2\theta },\nonumber\\
 \label{eq:law_diffrot}
\end{eqnarray}
where $A_c$ and $\Omega_c$ are constants.
Once the metric functions are obtained,
we can derive the angular velocity as a function of
$r$ and $\theta$,
which should be compared with the profile derived directly
from the equilibrium configuration.
Note that this relation is reduced to
a cylindrical rotation law in the non-relativistic limit.

As mentioned repeatedly, our new formulation 
can accommodate any EOS,
which may depend not only on density and entropy 
but also on other quantities
such as electron fraction and mass fractions of various nuclei.
In this paper, we employ a polytropic type of EOS for simplicity:
\begin{eqnarray}
 P(\rho, s(r,\theta)) = K(r,\theta) \rho^{1+\frac{1}{N}},
  \label{eq:eos}
\end{eqnarray}
where $N$ is the polytropic index;
note that $K$, which is normally a constant and is a function of
entropy, is assumed here to be a function of $r$ and $\theta$ as
\begin{eqnarray}
 K(r,\theta) \equiv \hat{K}\left(K_0
			+\epsilon_1\frac{r^2}{R_e^2}\sin^2\theta
			+\epsilon_2\frac{r^2}{R_e^2}\cos^2\theta
			+\epsilon_3\frac{r^2-R_e^2}{R_e^2}
		       \right),
  \label{eq:eosK}
\end{eqnarray}
where $R_e$ is the equatorial surface radius of the star;
$K_0, \epsilon_1, \epsilon_2, \epsilon_3$ are constants.
The polytropic index is set to $N=1$ in the following.
For the stationary models, we set $K_0=1$ as well.
Note that we use the geometrical unit~$c=G=1$
and put $\hat{K}=1$
in actual calculations following~(\cite{Cook:1992}).

It should be apparent that
the baroclinicity is introduced by the non-constancy of $K$
in this EOS. In fact
we set all $\epsilon$'s to $0$ for the barotropic case above.
In the baroclinic case, we solve the Einstein and Euler equations
keeping the specific angular momentum attached to grid points
 the same as that of the reference model.
We have to update the value of $K$
at each grid point according to its current position
 and Eq.~\eqref{eq:eos}.

\begin{figure*}
 \begin{tabular}{cc}
  \includegraphics[width=7.cm]{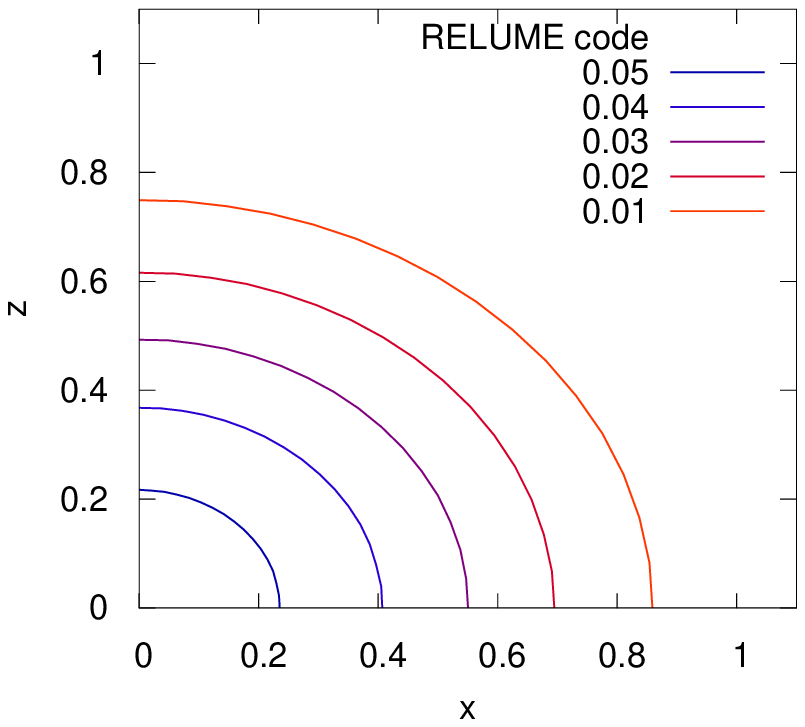} &
  \includegraphics[width=7.cm]{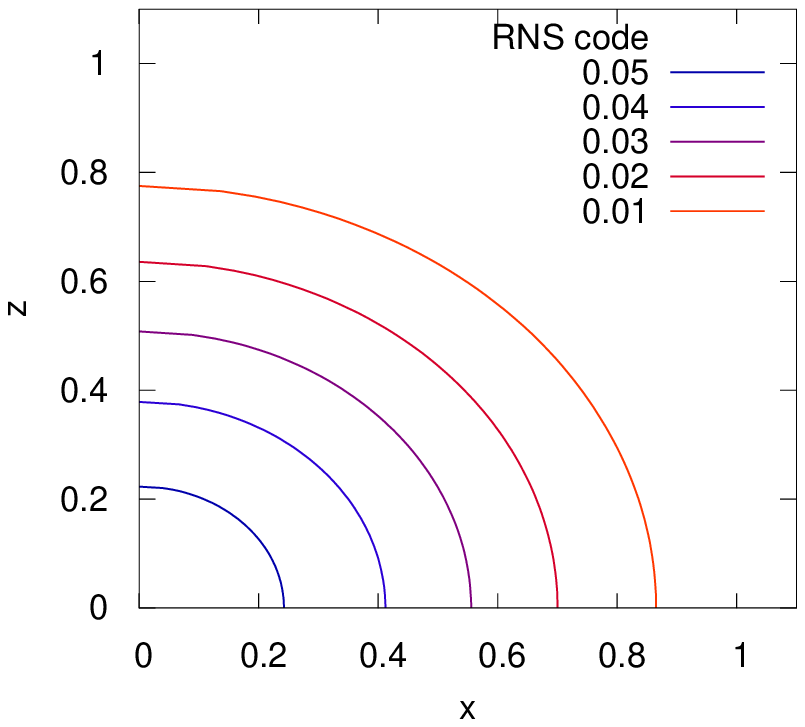}\\
  (a) & (b)
 \end{tabular}
 \caption{Density contour of the uniformly rotatng star 
 computed (a) by our new code
 and (b) by RNS code.}
\label{fig:comparison}
\end{figure*}
\begin{table*}
\centering
\begin{tabular}{cccccc}
 Code & $N_r\times N_{\theta}$ & $\rho_c$ & $r_p/r_e$ & $M$ & $M_b$\\\hline
 our code & $32\times 17$ & $0.0562$ & $0.874$ & $0.0958$ & $0.100$\\
 RNS code & $65\times 129$ & $0.0562$ & $0.870$ & $0.0967$ & $0.101$
\\\hline
\end{tabular}
 \caption{Parameters for comparison with the public RNS code by \citet{Stergioulas1995}.}
 \label{table:comparison}
\end{table*}

\subsection{Evolutionary models}
In its real evolution, the PNS cools
down through neutrino emissions
and experiences a sequence of quasi-equilibrium
during a period of about a minute.
If it rotates as normally expected,
these are rotational equilibria with different thermal
and lepton contents but with the same specific
angular momentum if the angular momentum transfer is negligible.
To understand such secular evolutions of PNS,
it is necessary to construct a series of configurations
of rotational equilibria and to compute
the neutrino transfer on top of them.
The main aim of this paper is to
provide a new numerical tool to treat the first step,
i.e., the building of a sequence of rotational equilibria that
are supposed to represent the secular evolution
of the rotating PNS.
The purpose of the models considered here is to
demonstrate the capability of our new Lagrangian formulation
in full general relativity with some mock evolutions.
The application of the method to more realistic evolutions is currently underway and will be presented elsewhere in the near future.

For the mock evolutions,
we consider the following three scenarios.
We stress that in our formulation we need to give an
angular velocity distribution 
only at the beginning of each sequence and
it is automatically derived thereafter
during the evolution\footnote{We need to specify the specific angular momentum of accreting matter in scenario~(iii).}.


\begin{enumerate}
 \item \underline{Cooling model}:
The first model is intended to mimic 
       a cooling evolution by the neutrino emission,
in which the PNS shrinks as the thermal energy is carried away by
neutrinos.
       In realistic simulations~(\citet{Pons:1998mm,Villain:2003ey}), we need to calculate the entropy~(and electron fraction) evolution with the neutrino transfer.
       For the demonstrative purpose here, 
       it suffices to prescribe the time-dependence of
$K$ in the Polytropic EOS by hand so that
it should mimic the cooling.
 \item \underline{Wind model}:
       The second model is to mimic the evolution
of a rotating star via the mass-loss from the surface as a wind.
The neutrino-driven wind from the PNS has been considered in~\citet{Meyer:1992zz,Witti:1992fn,Woosley:1994ux,Otsuki:1999kb,Sumiyoshi:1999rh,Terasawa:2001wn,Wanajo:2001pu,Panov:2008tr}.
 Such a stellar wind carries not only mass but also angular momentum.
In this toy model,
we emulate the mass-loss process by
taking a certain fraction of mass and angular momentum
away from the grid points near the stellar surface
at a certain rate.
 \item \underline{Accretion model}:
       In the third model, we consider the evolution
induced by the accretion of matter,
the process opposite to the one in model~(ii).
       Recent core-collapse supernova simulations~(e.g.~\citet{Muller:2018utr,Burrows2019,Nakamura:2019snn,Janka:2021deg}) indicate that
asymmetric accretion flows continue to exist
for a few seconds after the revival of a stalled shock wave
 and may affect the angular momentum of PNS
~(\citet{Blondin2007,Fernandez2010,Wongwathanarat:2012zp,Guilet:2013bxa,Kazeroni:2017fup}).
 In this model, we simply increase the masses
and angular momenta of the grid points near the surface
at a certain rate, to consider the spin-up evolution of PNS.
\end{enumerate}

%
\begin{table*}
\centering
\begin{tabular}{cccccccc}
 Model & Number of Grids($N_r\times N_{\theta}$) 
 & $R_p/R_e$ & $M_b$ & $M$ & $J$ & $T/|W|$ & GRV \\\hline
Reference & $32\times 17$ & $0.893$ & $0.0999$ & $0.0961$ & $3.91\times 10^{-3}$
 & $6.02\times 10^{-2}$ & $0.976\times 10^{-3}$ \\
Middle-Resolution & $16\times 9$ & $0.891$ & $0.0999$ & $0.0958$ & $3.77\times 10^{-3}$
 & $5.61\times 10^{-2}$ & $4.51\times 10^{-2}$ \\
Low-Resolution & $8\times 5$ & $0.885$ & $0.0986$ & $0.0950$ & $1.93\times 10^{-3}$
 & $3.66\times 10^{-2}$ & $1.22\times 10^{-1}$ \\
EOS-a  & $32\times 17$ & $0.874$ & $0.0994$ & $0.0946$ & $3.95\times 10^{-3}$
 & $5.14\times 10^{-2}$ & $2.07\times 10^{-3}$ \\
EOS-b  & $32\times 17$ & $0.911$ & $0.100$ & $0.0956$ & $3.87\times 10^{-3}$
 & $5.34\times 10^{-2}$ & $3.78\times 10^{-3}$ \\
EOS-c  & $32\times 17$ & $0.894$ & $0.0999$ & $0.0954$ & $3.91\times 10^{-3}$
 & $5.32\times 10^{-2}$ & $2.91\times 10^{-3}$ \\
Middle-Rotation  & $32\times 17$ & $0.812$ & $0.0997$ & $0.0951$ & $5.79\times 10^{-3}$
 & $9.88\times 10^{-2}$ & $6.75\times 10^{-3}$ \\
Cooling  & $32\times 17$ & $0.895$ & $0.0989$ & $0.0955$ & $3.85\times 10^{-3}$
 & $6.47\times 10^{-2}$ & $8.02\times 10^{-3}$ \\
Wind  & $32\times 17$ & $0.890$ & $0.0980$ & $0.0943$ & $3.80\times 10^{-3}$
 & $5.90\times 10^{-2}$ & $7.27\times 10^{-3}$ \\
Accretion  & $32\times 17$ & $0.895$ & $0.102$ & $0.0981$ & $4.22\times 10^{-3}$
 & $6.71\times 10^{-2}$ & $4.74\times 10^{-3}$
\end{tabular}
 \caption{
 List of models:
 The first column shows the model name.
 The second corresponds to the number of grid points,
 followed by the ratio of the polar radius to the equatorial radius of the star, the baryon mass, the gravitational mass, the total angular momentum, $T/|W|$, and relativistic Virial relation~(GRV2)
 in~\citet{Nozawa1998}. 
 }
 \label{table:model}
\end{table*}

\section{Results}\label{sec:result}
In this section, 
starting from the comparison of our solution with
that by the public RNS code,
we show barotropic stationary models and baroclinic stationary models, followed by the three mock evolutionary models, namely, cooling, wind, and accretion models.

\subsection{Uniformly rotating stars}
Fortunately, it is possible to directly compare
our result in the Lagrange formulation
with the solution given by the well-known public RNS code for uniformly-rotating stars~(\cite{Stergioulas1995}).
In Fig.~\ref{fig:comparison},
we show the density contour of a uniformly rotating star
(a) by our new code in the Lagrangian formulation 
and (b) by the RNS code in the literature.
Since the input parameters to construct equilibrium configurations are different in the Eurelian and Lagrangian formulations,
we finetune those parameters to obtain almost the same rotating star as possible as we can, 
which the global quantities are compared in Table~\ref{table:comparison}.
Note also that the relativistic Virial relation of our solution
is $8.94\times 10^{-3}$,
which indicates the accuracy of solutions~(\cite{Nozawa1998}).

\subsection{Stationary models with barotropic EOS}
\begin{figure}
 \includegraphics[width=8.5cm]{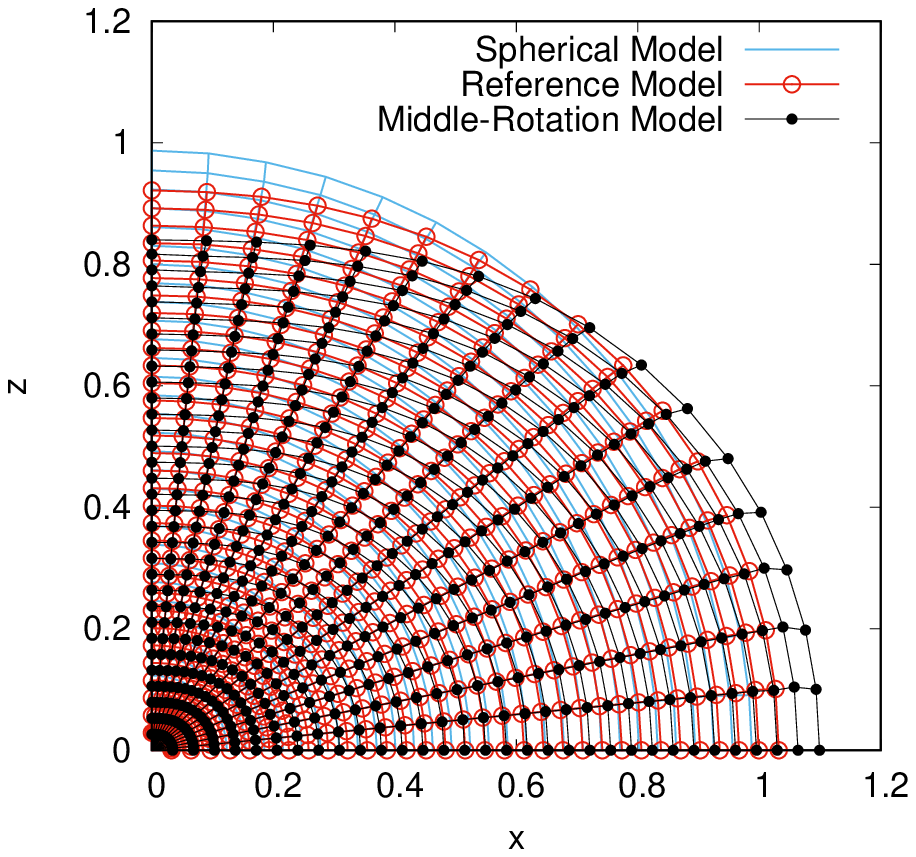}
 \caption{Stationary barotropic rotating stars
 with the differential rotation law.
 The shape of the reference model with the red circles
 is compared with those of the spherical model with the green lines
 and the middle-rotation model with the black circles.
 }
 \label{fig:diffrot}
\end{figure}
\begin{figure}
 \begin{tabular}{c}
  \includegraphics[width=8.5cm]{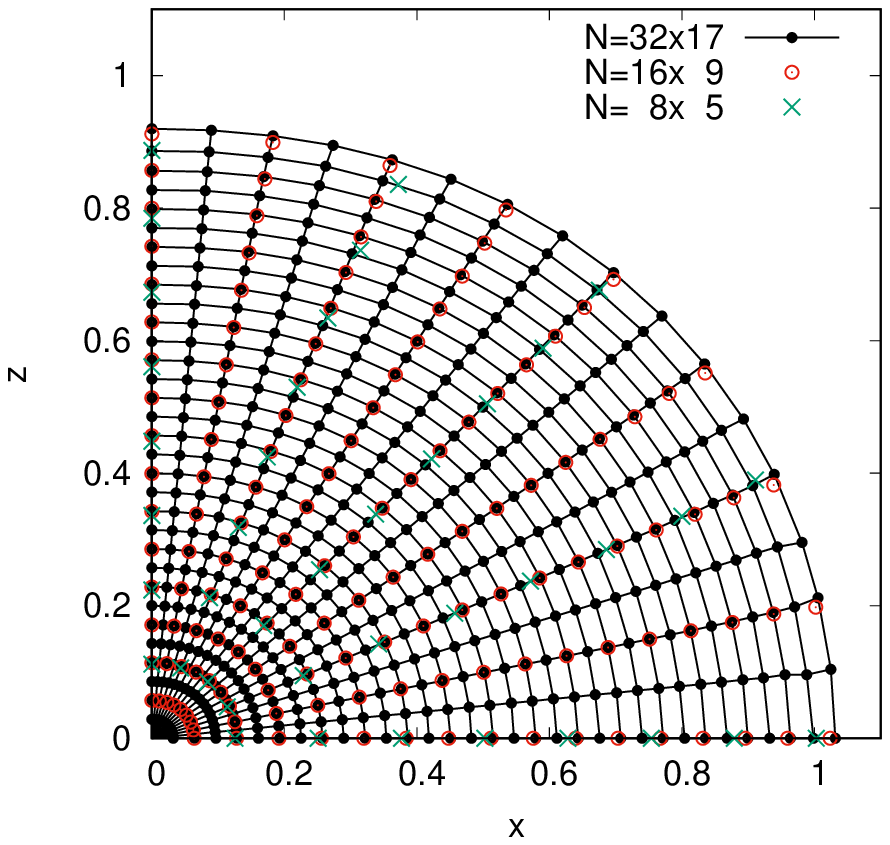}\\
  (a)
\\  \includegraphics[width=8.5cm]{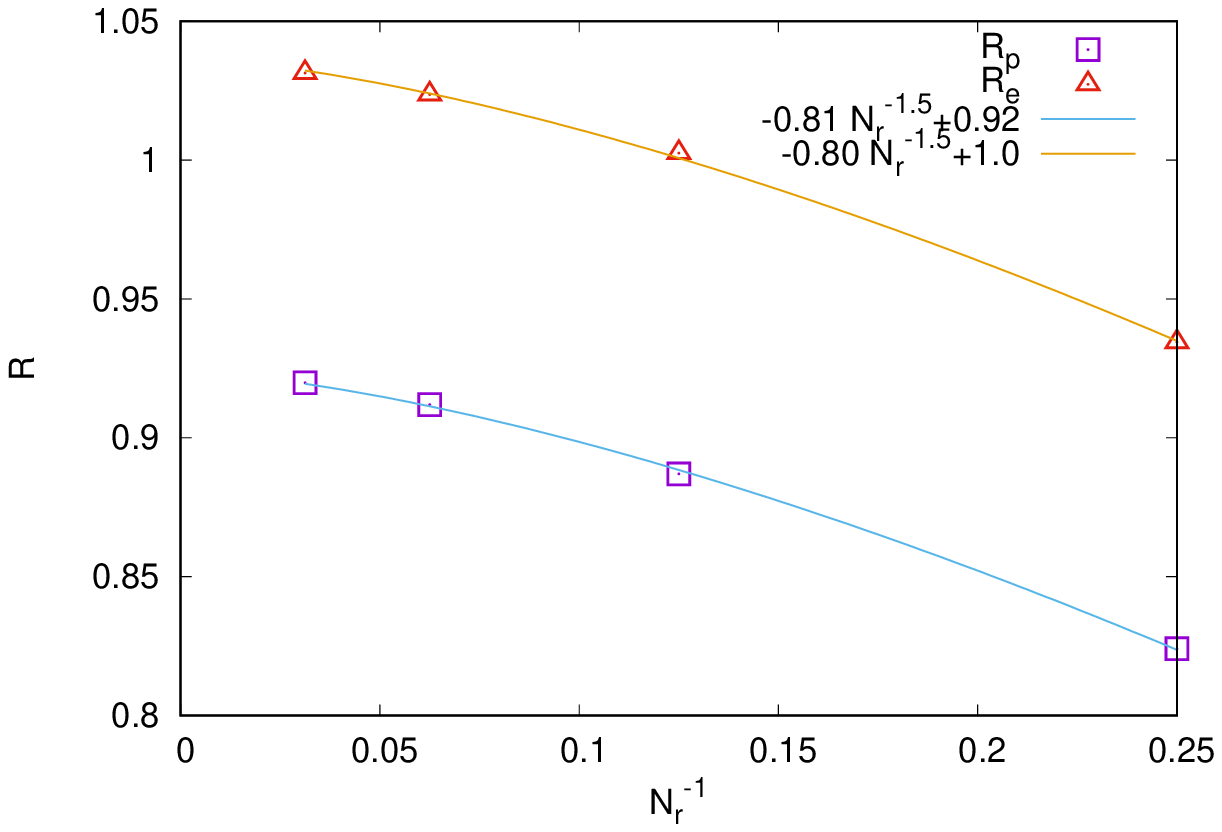}\\
  (b)
 \end{tabular}
 \caption{(a) Shape of the rotating stars 
 in the reference model with different
 resolutions. The lowest, middle and highest solutions are
 displayed with the green crosses,
 the red open circles, and the black filled circles.
 (b) Surface radii as a function of the number of radial meshes.
 The polar radii are shown by the purple squares,
 while the equatorial radii are shown by the red triangles,
 which are fitted by $R=\alpha N_r^{-\beta} +\gamma$ curves.}
 \label{fig:resolution}
\end{figure}
We first focus on the barotropic rotating stars with $\epsilon_1=\epsilon_2=\epsilon_3=0$ and the EOS described in Eq.~\eqref{eq:eos}.
As the angular velocity is faster,
the star is oblater.
In Fig.~\ref{fig:diffrot}, 
the spherical and middle-rotation models are compared 
with the reference model.
Furthermore, the resolution dependence is investigated in Fig.~\ref{fig:resolution}.
The solution converges as the resolution increases.
As shown in Fig.~\ref{fig:resolution}~(b),
the convergence order is 1.5, which is 
expected because we employ the first-order FE scheme to calculate the baryon density while we use the second-order FE scheme to evaluate the derivatives.
In this work, we adopt the model with $N_r\times N_{\theta}=32\times 17$ as the reference model for other computations.
Note that our solutions with the resolution~$(N_r\times N_{\theta}=32\times 17)$
 typically has the Virial relation of $\mathcal{O}(10^{-3})$
 in Table~\ref{table:model}.

\subsection{Stationary models with baroclinic EOS}
\begin{figure*}
\begin{tabular}{cc}
 \includegraphics[width=8.5cm]{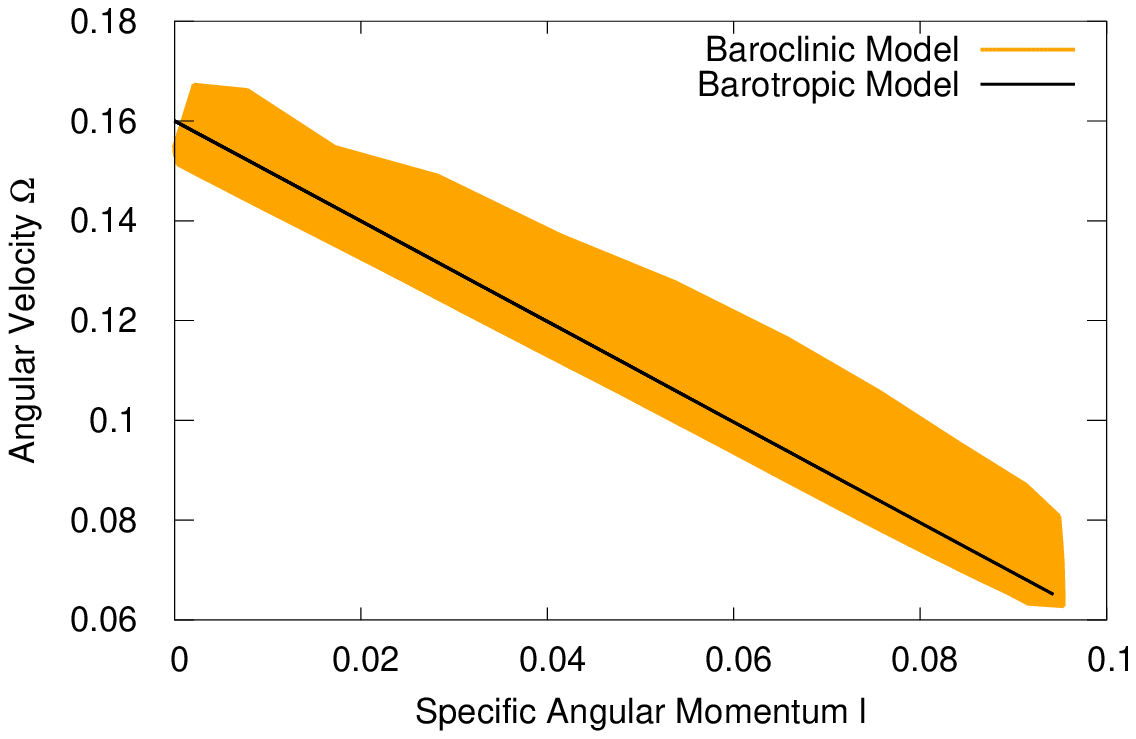} &
 \includegraphics[width=8.5cm]{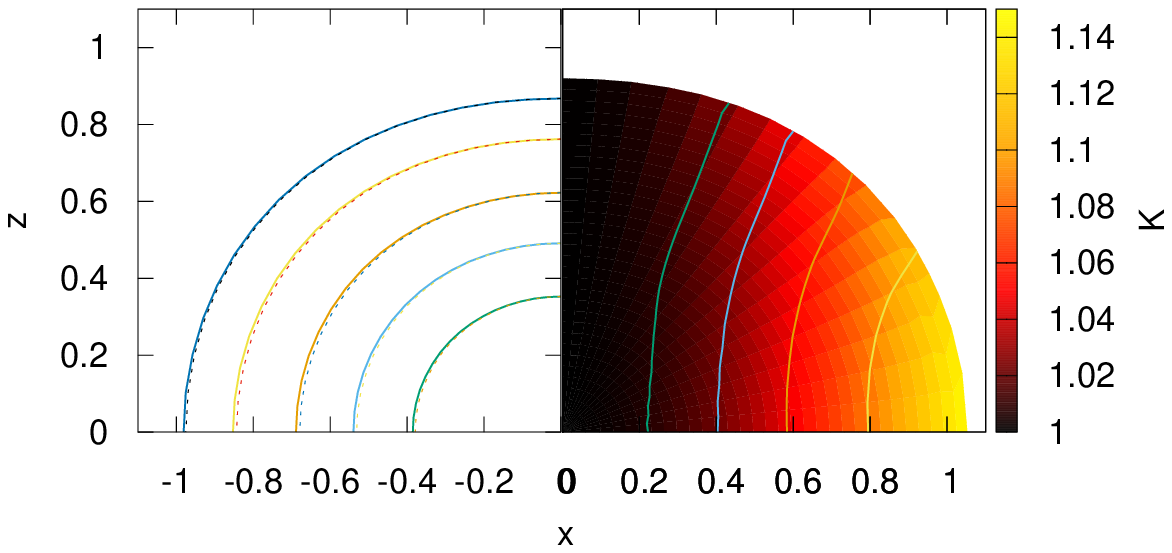}\\
 (a) $\Omega$ and $\ell$ of the model EOS-a &
     \begin{minipage}{0.5\textwidth}
      (b) Model EOS-a. (Left) Contours of pressure and density
      and (Right) color map of $K$
      and contour of $\Omega$
     \end{minipage} \\
 \includegraphics[width=8.5cm]{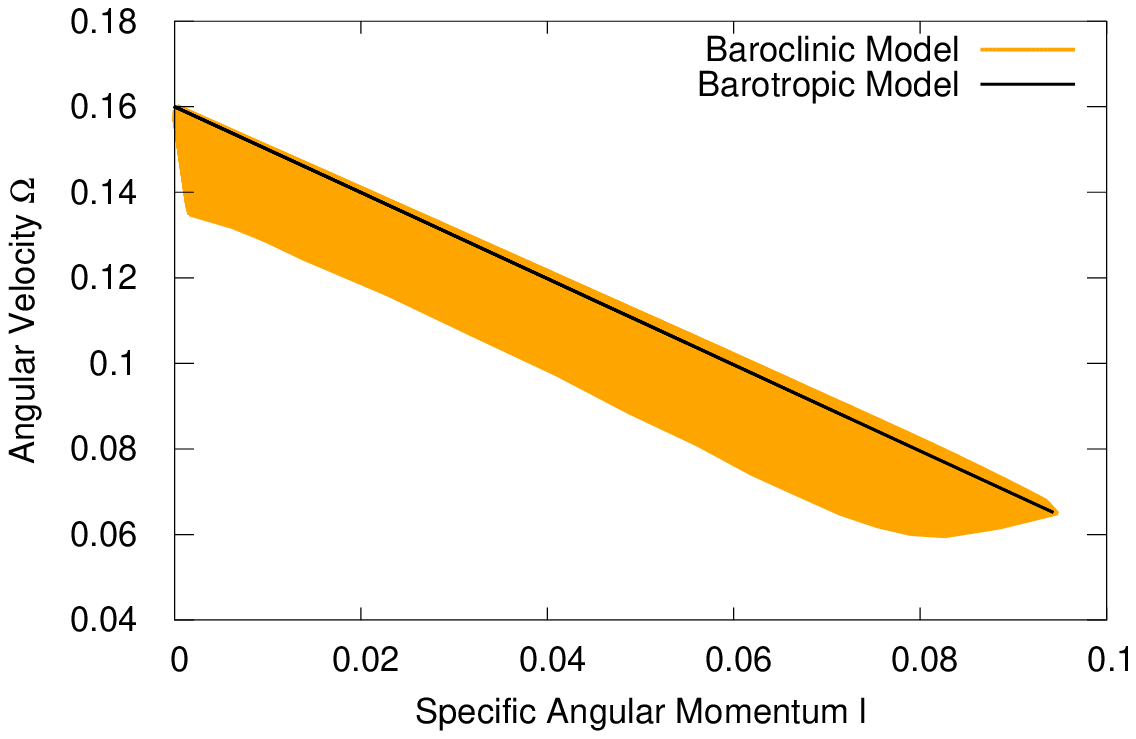} &
 \includegraphics[width=8.5cm]{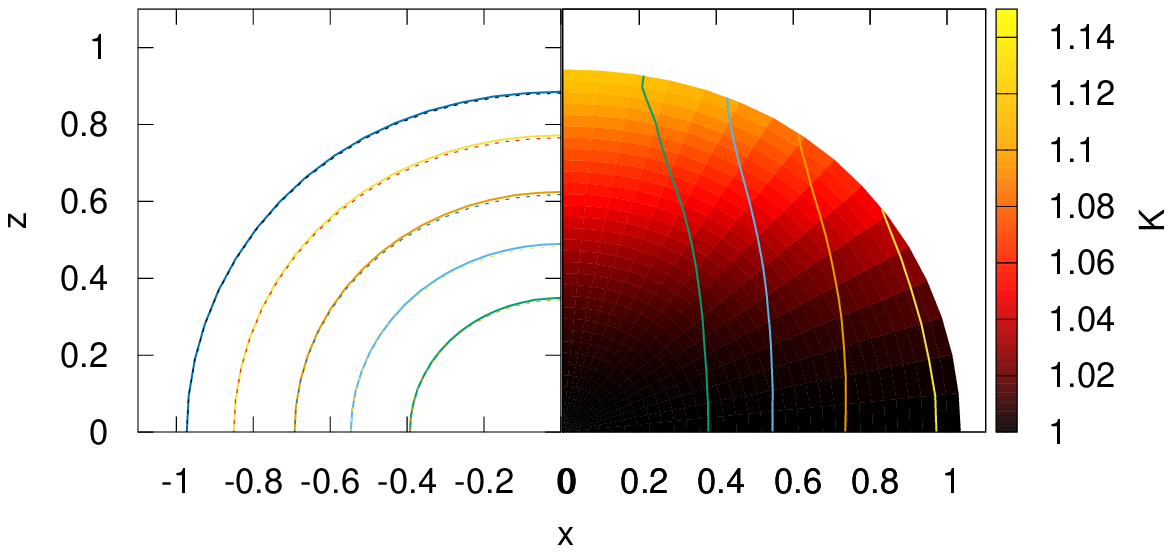} \\
 (c) $\Omega$ and $\ell$ of the model EOS-b &
     \begin{minipage}{0.5\textwidth}
      (d) Model EOS-b. (Left) Contours of pressure and density
      and (Right) color map of $K$
      and contour of $\Omega$
     \end{minipage} \\
 \includegraphics[width=8.5cm]{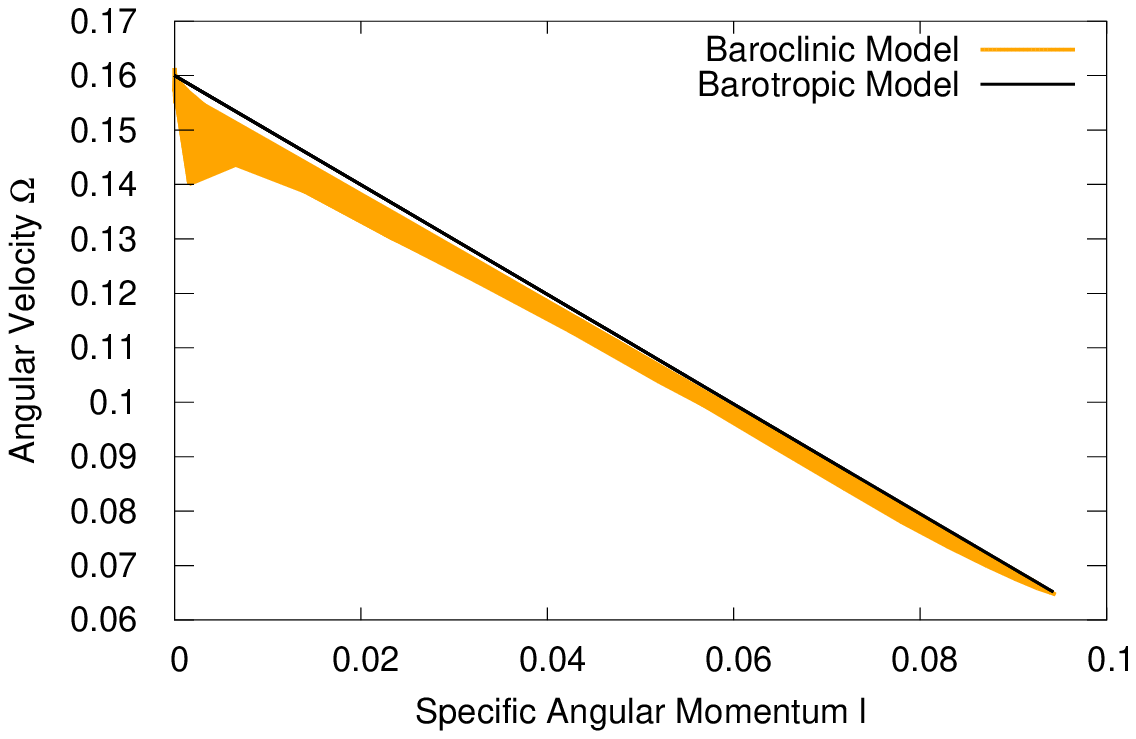} &
 \includegraphics[width=8.5cm]{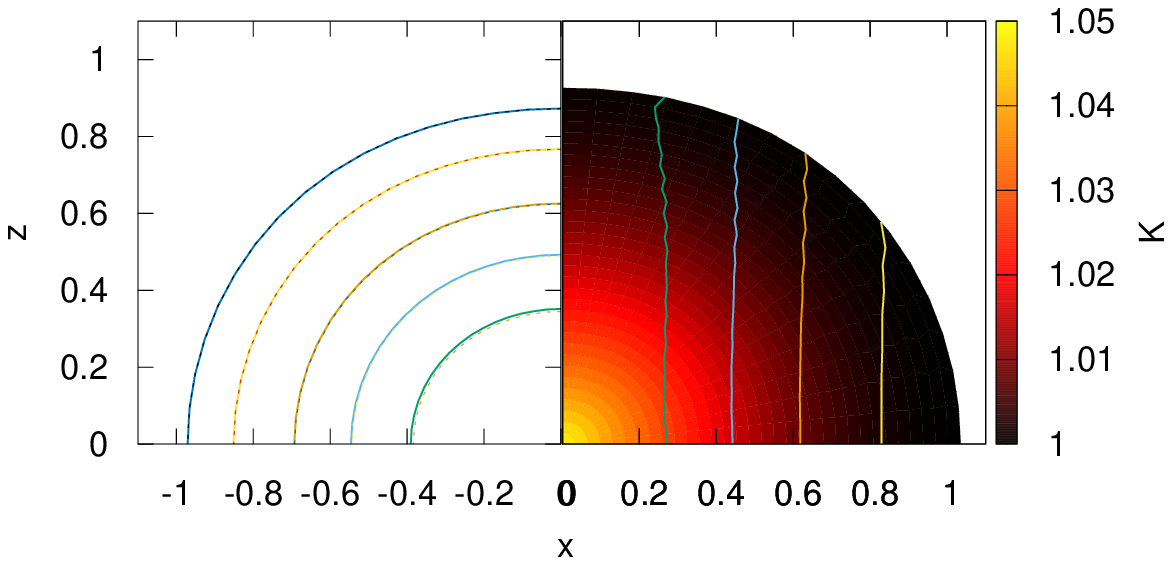} \\
 (e) $\Omega$ and $\ell$ of the model EOS-c &
     \begin{minipage}{0.5\textwidth}
      (f) Model EOS-c (Left) Contours of pressure and density
      and (Right) color map of $K$
      and contour of $\Omega$ 
     \end{minipage}
\end{tabular}
 \caption{Baroclinic rotating stars in general relativity.  The top, middle and bottom panels are the results of
the model EOS-a ($\epsilon_1=0.15, \epsilon_2=\epsilon_3=0$),
EOS-b ($\epsilon_2=0.15, \epsilon_3=\epsilon_1=0$)
 and EOS-c ($\epsilon_3=0.05, \epsilon_1=\epsilon_2=0$)
 in the entropy configuration, i.e., Eq.~\eqref{eq:eosK}, respectively.
 The left column shows the angular velocity as a function of the angular momentum of baroclinic rotating stars 
and that of barotropic ones shown by the black line.
The right column displays the misalignment between the density~(dashed)
 and pressure~(solid) contours in the left panel, which directly indicates the baroclinicity.
In the right panel of the right column,
the color map of $K(r,\theta)$ is shown as well as
the contour of angular velocity.
 }
 \label{fig:baroclinic}
\end{figure*}

Next, we investigate the effect of the baroclinicity on the equilibrium configuration of rotating stars.
In Fig.~\ref{fig:baroclinic}, we observe the baroclinic feature,
which is the misalignment between the pressure and density gradients
as Figs.~\ref{fig:baroclinic}~(b), (d) and (f).
To clearly see the difference between the barotropic and baroclinic models, we show the angular velocity as a function of the specific angular momentum in 
Figs.~\ref{fig:baroclinic}~(a), (c) and (e) as shown in~\cite{Camelio:2019rsz}.
Isentropic rotating stars in general relativity
satisfy the condition that 
is expressed as the black solid line by $\ell = \ell(\Omega)$
from Eqs.~\eqref{eq:Fomegaell} and~\eqref{eq:law_diffrot}.
Although a type of shellular rotation has been considered
as the initial rotation in core-collapse
supernova simulations~(e.g.~\citet{Yamada:1994,Harada:2018ubo,Iwakami:2021pwo}), the self-consistent stationary solutions 
with baroclinicity have been
obtained only in the Newtonian gravity
so far~(\citet{Roxburgh2006,Fujisawa:2015}).
We first obtain such a self-consistent stationary solution 
with baroclinicity in general relativity 
in Fig.~\ref{fig:baroclinic}~(d).
Since the model~(EOS-c) does not satisfy
the H{\o}iland criterion in the Newtonian limit
and may be dynamically unstable~(\cite{Tassoul:1978}),
we leave the detailed investigation on the dynamical stability 
for baroclinic rotating stars in the future study.

\subsection{Cooling of rotating stars}
\begin{figure}
 \begin{tabular}{c}
  \includegraphics[width=8.5cm]{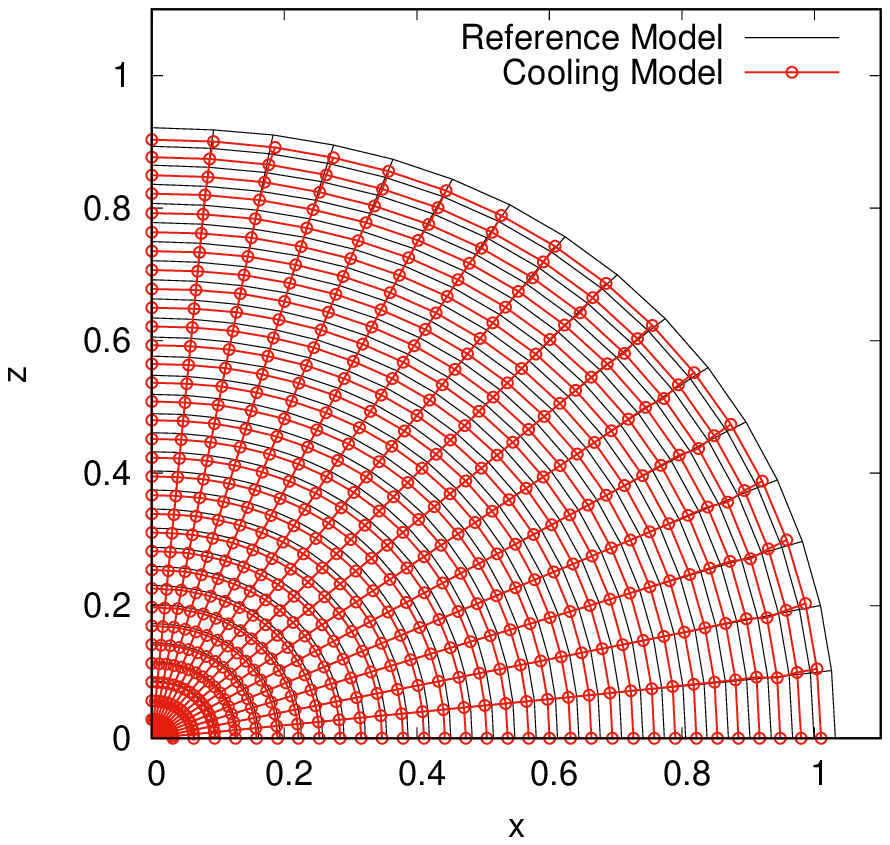}\\
  (a)\\
  \includegraphics[width=8.5cm]{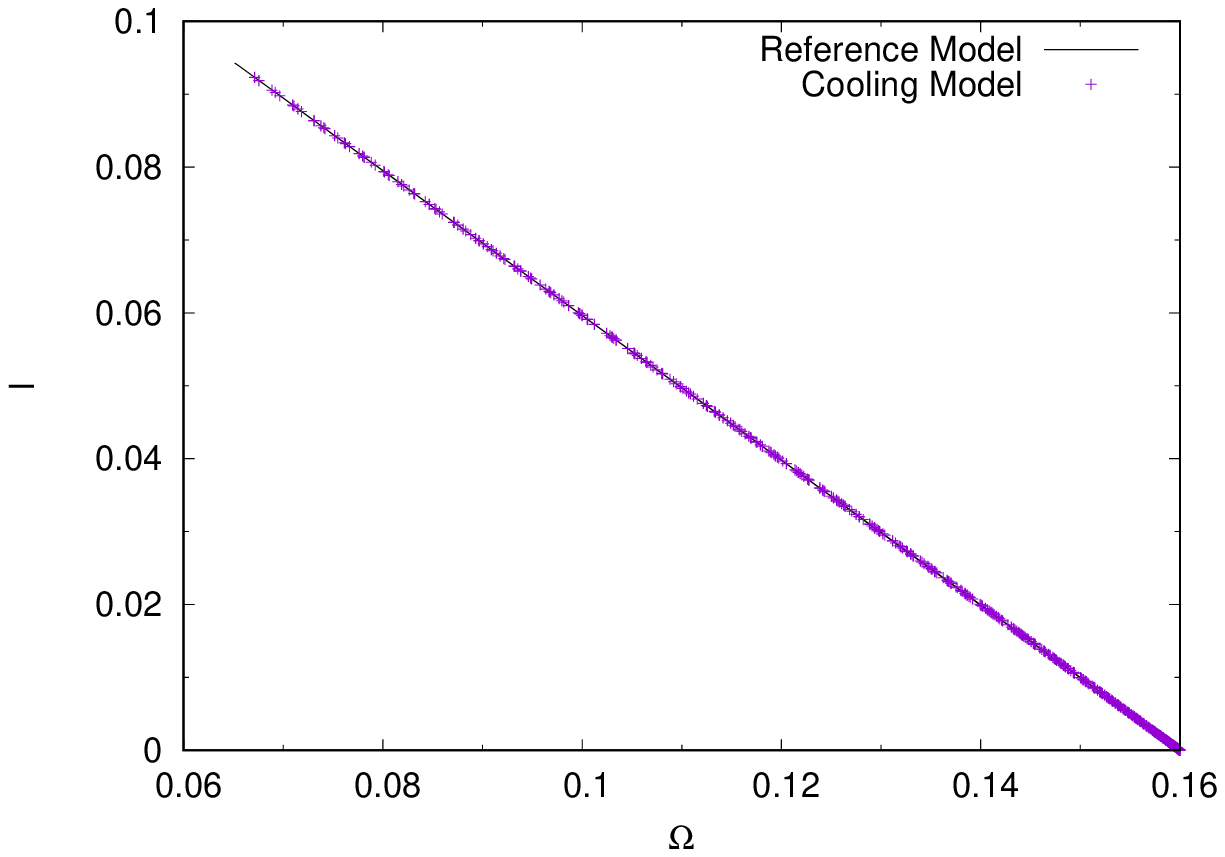}\\
  (b)
 \end{tabular}
 \caption{(a) Rotating stars for cooling model~($K_0=0.95$)
 compared to the reference model~($K_0=1$)
 (b) Angular velocity as a function of the specific angular momentum
 of the reference and cooling models
 }
 \label{fig:coolingtest}
\end{figure}

As an evolution test by cooling,
 we vary the polytropic constant~$K_0$
keeping the other parameters constant.
In Fig.~\ref{fig:coolingtest}~(a),
we compare the cooling model~($K_0=0.95$)
 with the reference model~($K_0=1$).
Rotating stars shrink for nonlinear balance equations between the pressure and the gravity to be satisfied.
We emphasize that we do not impose any rotational law explicitly but keep the specific angular momentum~$j_{\vphi}$ of each FE.
As mentioned, barotropic rotating stars satisfy
the condition~$\ell=\ell(\Omega)$ between $\ell$ and $\Omega$
 for any equilibria configurations.
In fact, Fig.~\ref{fig:coolingtest}~(b) shows that the angular velocity of the cooling model as a function of the specific angular momentum is still expressed as the line, which is highly non-trivial in this Lagrangian formulation.

\subsection{Wind and accretion models}
\begin{figure}
 \begin{tabular}{c}
  \includegraphics[width=8.5cm]{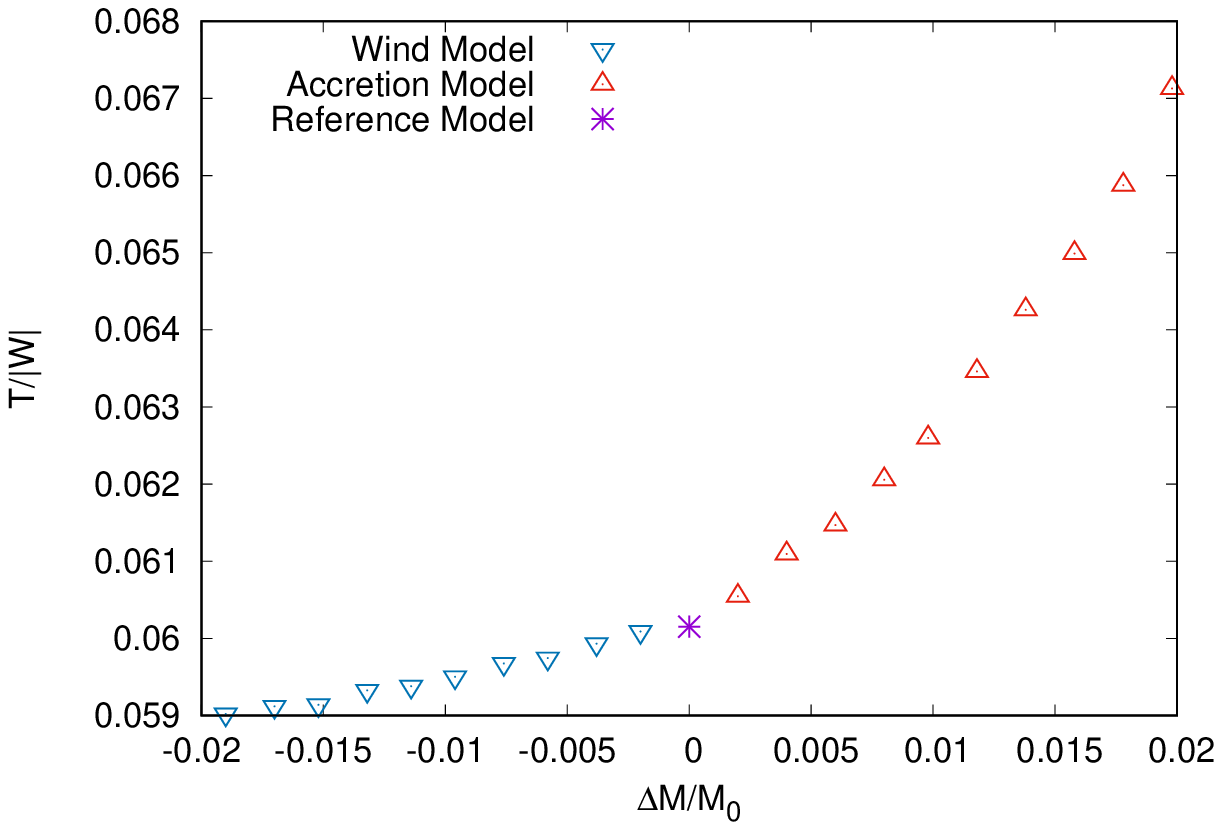}
 \end{tabular}
 \caption{T/|W| as a function of the mass variation
  by the wind and accretion.
 }
 \label{fig:masschange}
\end{figure}

For the wind model, we assume
the mass-loss from the outer layer of rotating stars.
Specifically, we decrease the specific masses of the outer two layers by hand as a toy model and compute a new equilibrium configuration.
For the accretion model, on the other hand,
we assume the accreting matter that is composed of both the $5\%$ mass 
and angular momentum of
outer layers of rotating stars for simplicity.
By the mass-loss or accretion, the total mass
and total angular momentum are expected to change accordingly.
In Fig.~\ref{fig:masschange},
we show $T/|W|$ as a function of the mass change.



%

\section{Conclusions}\label{sec:conclusion}
We have proposed a new scheme to 
construct the equilibrium configuration
of general relativistic rotating stars in the Lagrangian
formulation,
which maintains the mass and angular momentum automatically
when no angular momentum transfer exists, in contrast to the Eulerian
formulation.
We employed the traditional iterative scheme~(\cite{Hachisu:1986})
to solve the whole system derived from discretizing the Einstein and Euler equations on the finite-elements grid,
 which consists of two new schemes of our own devising:
for the Einstein equation, the W4IX method in App.~\ref{app:W4IX} is applied and
 for the Euler equation, the slice-shooting scheme is adopted in
App.~\ref{app:sliceshooting}.

Our formulation does need an angular velocity profile only at initial
and employ any equation of state.
It enables us to find the evolutionary sequence
by keeping the mass, specific angular momentum, 
specific entropy, electron fraction and mass fraction of
various nuclei.
In order to demonstrate the capability of our new formulation,
we first compare our result of uniformly-rotating stars
with that given by the public RNS code~(\cite{Stergioulas1995}).
Next, we present the stationary rotating stars with 
the barotropic and baroclinic equations of state.
Finally, we consider the mock evolutionary sequences 
for cooling, wind and accretion as toy models
and show those results.


In this paper, we adopt simple models for finding the evolutionary sequence, since our focus is on the way of constructing the equilibrium configuration in the Lagrangian formulation.
The application of our new method to more realistic situations 
is currently underway and will be presented in the near future.


\section*{Acknowledgements}

We would like to thank K-i. Maeda for helpful comments.
This work was supported by JSPS KAKENHI Grant Number 
20K03951, 20K03953, 20K14512, 20H04728, 20H04742,
and by Waseda University Grant for Special Research Projects(Project Number: 2019C-640).
\section*{Data Availability}
The data underlying this paper will be available from the corresponding author on reasonable request.



\bibliographystyle{mnras}
\bibliography{grstar} 

\begin{thebibliography}{}
\makeatletter
\relax
\def\mn@urlcharsother{\let\do\@makeother \do\$\do\&\do\#\do\^\do\_\do\%\do\~}
\def\mn@doi{\begingroup\mn@urlcharsother \@ifnextchar [ {\mn@doi@}
  {\mn@doi@[]}}
\def\mn@doi@[#1]#2{\def\@tempa{#1}\ifx\@tempa\@empty \href
  {http://dx.doi.org/#2} {doi:#2}\else \href {http://dx.doi.org/#2} {#1}\fi
  \endgroup}
\def\mn@eprint#1#2{\mn@eprint@#1:#2::\@nil}
\def\mn@eprint@arXiv#1{\href {http://arxiv.org/abs/#1} {{\tt arXiv:#1}}}
\def\mn@eprint@dblp#1{\href {http://dblp.uni-trier.de/rec/bibtex/#1.xml}
  {dblp:#1}}
\def\mn@eprint@#1:#2:#3:#4\@nil{\def\@tempa {#1}\def\@tempb {#2}\def\@tempc
  {#3}\ifx \@tempc \@empty \let \@tempc \@tempb \let \@tempb \@tempa \fi \ifx
  \@tempb \@empty \def\@tempb {arXiv}\fi \@ifundefined
  {mn@eprint@\@tempb}{\@tempb:\@tempc}{\expandafter \expandafter \csname
  mn@eprint@\@tempb\endcsname \expandafter{\@tempc}}}

\bibitem[\protect\citeauthoryear{Abbott et~al.,}{Abbott
  et~al.}{2016}]{PhysRevLett.116.061102}
Abbott B.~P.,  et~al., 2016, \mn@doi [Phys. Rev. Lett.]
  {10.1103/PhysRevLett.116.061102}, 116, 061102

\bibitem[\protect\citeauthoryear{Abbott et~al.,}{Abbott
  et~al.}{2017}]{PhysRevLett.119.161101}
Abbott B.~P.,  et~al., 2017, \mn@doi [Phys. Rev. Lett.]
  {10.1103/PhysRevLett.119.161101}, 119, 161101

\bibitem[\protect\citeauthoryear{Abbott et~al.,}{Abbott
  et~al.}{2021}]{Abbott_2021}
Abbott R.,  et~al., 2021, \mn@doi [The Astrophysical Journal Letters]
  {10.3847/2041-8213/ac082e}, 915, L5

\bibitem[\protect\citeauthoryear{{Ansorg}, {Kleinw{\"a}chter}  \&
  {Meinel}}{{Ansorg} et~al.}{2002}]{Ansorg:2002}
{Ansorg} M.,  {Kleinw{\"a}chter} A.,   {Meinel} R.,  2002, \mn@doi [\aap]
  {10.1051/0004-6361:20011643}, \href
  {https://ui.adsabs.harvard.edu/abs/2002A&A...381L..49A} {381, L49}

\bibitem[\protect\citeauthoryear{Bathe}{Bathe}{2006}]{bathe2006finite}
Bathe K.-J.,  2006, Finite element procedures.
Klaus-Jurgen Bathe

\bibitem[\protect\citeauthoryear{Birkl, Stergioulas  \& Muller}{Birkl
  et~al.}{2011}]{Birkl:2010hc}
Birkl R.,  Stergioulas N.,   Muller E.,  2011, \mn@doi [Phys. Rev. D]
  {10.1103/PhysRevD.84.023003}, 84, 023003

\bibitem[\protect\citeauthoryear{{Blondin} \& {Mezzacappa}}{{Blondin} \&
  {Mezzacappa}}{2007}]{Blondin2007}
{Blondin} J.~M.,  {Mezzacappa} A.,  2007, \mn@doi [\nat] {10.1038/nature05428},
  \href {https://ui.adsabs.harvard.edu/abs/2007Natur.445...58B} {445, 58}

\bibitem[\protect\citeauthoryear{{Bonazzola}, {Gourgoulhon}, {Salgado}  \&
  {Marck}}{{Bonazzola} et~al.}{1993}]{Bonazzola:1993}
{Bonazzola} S.,  {Gourgoulhon} E.,  {Salgado} M.,   {Marck} J.~A.,  1993, \aap,
  \href {https://ui.adsabs.harvard.edu/abs/1993A&A...278..421B} {278, 421}

\bibitem[\protect\citeauthoryear{Burrows \& Lattimer}{Burrows \&
  Lattimer}{1986}]{Burrows:1986me}
Burrows A.,  Lattimer J.~M.,  1986, \mn@doi [Astrophys. J.] {10.1086/164405},
  307, 178

\bibitem[\protect\citeauthoryear{Burrows \& Vartanyan}{Burrows \&
  Vartanyan}{2021}]{Burrows:2020qrp}
Burrows A.,  Vartanyan D.,  2021, \mn@doi [Nature]
  {10.1038/s41586-020-03059-w}, 589, 29

\bibitem[\protect\citeauthoryear{Burrows, Radice, Vartanyan, Nagakura, Skinner
  \& Dolence}{Burrows et~al.}{2019}]{Burrows2019}
Burrows A.,  Radice D.,  Vartanyan D.,  Nagakura H.,  Skinner M.~A.,   Dolence
  J.~C.,  2019, \mn@doi [Monthly Notices of the Royal Astronomical Society]
  {10.1093/mnras/stz3223}, 491, 2715

\bibitem[\protect\citeauthoryear{Burrows, Radice, Vartanyan, Nagakura, Skinner
  \& Dolence}{Burrows et~al.}{2020}]{Burrows:2019zce}
Burrows A.,  Radice D.,  Vartanyan D.,  Nagakura H.,  Skinner M.~A.,   Dolence
  J.,  2020, \mn@doi [Mon. Not. Roy. Astron. Soc.] {10.1093/mnras/stz3223},
  491, 2715

\bibitem[\protect\citeauthoryear{{Butterworth} \& {Ipser}}{{Butterworth} \&
  {Ipser}}{1976}]{Butterworth:1976}
{Butterworth} E.~M.,  {Ipser} J.~R.,  1976, \mn@doi [\apj] {10.1086/154163},
  \href {https://ui.adsabs.harvard.edu/abs/1976ApJ...204..200B} {204, 200}

\bibitem[\protect\citeauthoryear{Camelio, Gualtieri, Pons  \& Ferrari}{Camelio
  et~al.}{2016}]{Camelio:2016fan}
Camelio G.,  Gualtieri L.,  Pons J.~A.,   Ferrari V.,  2016, \mn@doi [Phys.
  Rev. D] {10.1103/PhysRevD.94.024008}, 94, 024008

\bibitem[\protect\citeauthoryear{Camelio, Dietrich, Marques  \&
  Rosswog}{Camelio et~al.}{2019}]{Camelio:2019rsz}
Camelio G.,  Dietrich T.,  Marques M.,   Rosswog S.,  2019, \mn@doi [Phys. Rev.
  D] {10.1103/PhysRevD.100.123001}, 100, 123001

\bibitem[\protect\citeauthoryear{{Cook}, {Shapiro}  \& {Teukolsky}}{{Cook}
  et~al.}{1992}]{Cook:1992}
{Cook} G.~B.,  {Shapiro} S.~L.,   {Teukolsky} S.~A.,  1992, \mn@doi [\apj]
  {10.1086/171849}, \href
  {https://ui.adsabs.harvard.edu/abs/1992ApJ...398..203C} {398, 203}

\bibitem[\protect\citeauthoryear{Cook, Shapiro  \& Teukolsky}{Cook
  et~al.}{1994}]{Cook:1993qj}
Cook G.~B.,  Shapiro S.~L.,   Teukolsky S.~A.,  1994, Astrophys. J., 422, 227

\bibitem[\protect\citeauthoryear{Diener, Rosswog  \& Torsello}{Diener
  et~al.}{2022}]{Diener:2022hui}
Diener P.,  Rosswog S.,   Torsello F.,  2022

\bibitem[\protect\citeauthoryear{{Espinosa Lara} \& {Rieutord}}{{Espinosa Lara}
  \& {Rieutord}}{2007}]{Espinosa-Lara2007}
{Espinosa Lara} F.,  {Rieutord} M.,  2007, \mn@doi [\aap]
  {10.1051/0004-6361:20077263}, \href
  {https://ui.adsabs.harvard.edu/abs/2007A&A...470.1013E} {470, 1013}

\bibitem[\protect\citeauthoryear{{Espinosa Lara} \& {Rieutord}}{{Espinosa Lara}
  \& {Rieutord}}{2013}]{Espinosa-Lara2013}
{Espinosa Lara} F.,  {Rieutord} M.,  2013, \mn@doi [\aap]
  {10.1051/0004-6361/201220844}, \href
  {https://ui.adsabs.harvard.edu/abs/2013A&A...552A..35E} {552, A35}

\bibitem[\protect\citeauthoryear{{Fern{\'a}ndez}}{{Fern{\'a}ndez}}{2010}]{Fernandez2010}
{Fern{\'a}ndez} R.,  2010, \mn@doi [\apj] {10.1088/0004-637X/725/2/1563}, \href
  {https://ui.adsabs.harvard.edu/abs/2010ApJ...725.1563F} {725, 1563}

\bibitem[\protect\citeauthoryear{Fujibayashi, Takahashi, Sekiguchi  \&
  Shibata}{Fujibayashi et~al.}{2021}]{Fujibayashi:2021wvv}
Fujibayashi S.,  Takahashi K.,  Sekiguchi Y.,   Shibata M.,  2021, ]
  {10.3847/1538-4357/ac10cb}

\bibitem[\protect\citeauthoryear{{Fujisawa}}{{Fujisawa}}{2015}]{Fujisawa:2015}
{Fujisawa} K.,  2015, \mn@doi [\mnras] {10.1093/mnras/stv2175}, \href
  {https://ui.adsabs.harvard.edu/abs/2015MNRAS.454.3060F} {454, 3060}

\bibitem[\protect\citeauthoryear{Fujisawa, Okawa, Yamamoto  \& Yamada}{Fujisawa
  et~al.}{2019}]{Fujisawa:2018dnh}
Fujisawa K.,  Okawa H.,  Yamamoto Y.,   Yamada S.,  2019, \mn@doi [Astrophys.
  J.] {10.3847/1538-4357/aaffdd}, 872, 155

\bibitem[\protect\citeauthoryear{Gourgoulhon, Bejger  \& Mancini}{Gourgoulhon
  et~al.}{2015}]{Gourgoulhon_2015}
Gourgoulhon E.,  Bejger M.,   Mancini M.,  2015, \mn@doi [Journal of Physics:
  Conference Series] {10.1088/1742-6596/600/1/012002}, 600, 012002

\bibitem[\protect\citeauthoryear{Goussard, Haensel  \& Zdunik}{Goussard
  et~al.}{1997}]{Goussard:1996dp}
Goussard J.-O.,  Haensel P.,   Zdunik J.~L.,  1997, Astron. Astrophys., 321,
  822

\bibitem[\protect\citeauthoryear{Goussard, Haensel  \& Zdunik}{Goussard
  et~al.}{1998}]{Goussard:1997bn}
Goussard J.~O.,  Haensel P.,   Zdunik J.~L.,  1998, Astron. Astrophys., 330,
  1005

\bibitem[\protect\citeauthoryear{Guilet \& Fern\'andez}{Guilet \&
  Fern\'andez}{2014}]{Guilet:2013bxa}
Guilet J.,  Fern\'andez R.,  2014, \mn@doi [Mon. Not. Roy. Astron. Soc.]
  {10.1093/mnras/stu718}, 441, 2782

\bibitem[\protect\citeauthoryear{{Hachisu}}{{Hachisu}}{1986}]{Hachisu:1986}
{Hachisu} I.,  1986, \mn@doi [\apjs] {10.1086/191121}, \href
  {https://ui.adsabs.harvard.edu/abs/1986ApJS...61..479H} {61, 479}

\bibitem[\protect\citeauthoryear{Harada, Nagakura, Iwakami, Okawa, Furusawa,
  Matsufuru, Sumiyoshi  \& Yamada}{Harada et~al.}{2019}]{Harada:2018ubo}
Harada A.,  Nagakura H.,  Iwakami W.,  Okawa H.,  Furusawa S.,  Matsufuru H.,
  Sumiyoshi K.,   Yamada S.,  2019, \mn@doi [Astrophys. J.]
  {10.3847/1538-4357/ab0203}, 872, 181

\bibitem[\protect\citeauthoryear{{Hartle}}{{Hartle}}{1967}]{Hartle:1967}
{Hartle} J.~B.,  1967, \mn@doi [\apj] {10.1086/149400}, \href
  {https://ui.adsabs.harvard.edu/abs/1967ApJ...150.1005H} {150, 1005}

\bibitem[\protect\citeauthoryear{Hirai, Nagakura, Okawa  \& Fujisawa}{Hirai
  et~al.}{2016}]{Hirai:2016ogc}
Hirai R.,  Nagakura H.,  Okawa H.,   Fujisawa K.,  2016, \mn@doi [Phys. Rev. D]
  {10.1103/PhysRevD.93.083006}, 93, 083006

\bibitem[\protect\citeauthoryear{Iwakami et~al.,}{Iwakami
  et~al.}{2021}]{Iwakami:2021pwo}
Iwakami W.,  et~al., 2021

\bibitem[\protect\citeauthoryear{Janka, Hanke, Hüdepohl, Marek, Müller  \&
  Obergaulinger}{Janka et~al.}{2012}]{Janka2012}
Janka H.-T.,  Hanke F.,  Hüdepohl L.,  Marek A.,  Müller B.,   Obergaulinger
  M.,  2012, \mn@doi [Progress of Theoretical and Experimental Physics]
  {10.1093/ptep/pts067}, 2012

\bibitem[\protect\citeauthoryear{Janka, Wongwathanarat  \& Kramer}{Janka
  et~al.}{2021}]{Janka:2021deg}
Janka H.-T.,  Wongwathanarat A.,   Kramer M.,  2021

\bibitem[\protect\citeauthoryear{Kazeroni, Guilet  \& Foglizzo}{Kazeroni
  et~al.}{2017}]{Kazeroni:2017fup}
Kazeroni R.,  Guilet J.,   Foglizzo T.,  2017, \mn@doi [Mon. Not. Roy. Astron.
  Soc.] {10.1093/mnras/stx1566}, 471, 914

\bibitem[\protect\citeauthoryear{Keil, Janka  \& Muller}{Keil
  et~al.}{1996}]{Keil:1996ab}
Keil W.,  Janka H.~T.,   Muller E.,  1996, \mn@doi [Astrophys. J. Lett.]
  {10.1086/310404}, 473, L111

\bibitem[\protect\citeauthoryear{{Komatsu}, {Eriguchi}  \& {Hachisu}}{{Komatsu}
  et~al.}{1989a}]{Komatsu:1989}
{Komatsu} H.,  {Eriguchi} Y.,   {Hachisu} I.,  1989a, \mn@doi [\mnras]
  {10.1093/mnras/237.2.355}, \href
  {https://ui.adsabs.harvard.edu/abs/1989MNRAS.237..355K} {237, 355}

\bibitem[\protect\citeauthoryear{{Komatsu}, {Eriguchi}  \& {Hachisu}}{{Komatsu}
  et~al.}{1989b}]{Komatsu1989b}
{Komatsu} H.,  {Eriguchi} Y.,   {Hachisu} I.,  1989b, \mn@doi [\mnras]
  {10.1093/mnras/239.1.153}, \href
  {https://ui.adsabs.harvard.edu/abs/1989MNRAS.239..153K} {239, 153}

\bibitem[\protect\citeauthoryear{Meyer, Mathews, Howard, Woosley  \&
  Hoffman}{Meyer et~al.}{1992}]{Meyer:1992zz}
Meyer B.~S.,  Mathews G.~J.,  Howard W.~M.,  Woosley S.~E.,   Hoffman R.~D.,
  1992, \mn@doi [Astrophys. J.] {10.1086/171957}, 399, 656

\bibitem[\protect\citeauthoryear{M\"uller et~al.,}{M\"uller
  et~al.}{2019}]{Muller:2018utr}
M\"uller B.,  et~al., 2019, \mn@doi [Mon. Not. Roy. Astron. Soc.]
  {10.1093/mnras/stz216}, 484, 3307

\bibitem[\protect\citeauthoryear{Nagakura et~al.,}{Nagakura
  et~al.}{2018}]{Nagakura:2017mnp}
Nagakura H.,  et~al., 2018, \mn@doi [Astrophys. J.] {10.3847/1538-4357/aaac29},
  854, 136

\bibitem[\protect\citeauthoryear{Nakamura, Takiwaki  \& Kotake}{Nakamura
  et~al.}{2019}]{Nakamura:2019snn}
Nakamura K.,  Takiwaki T.,   Kotake K.,  2019, \mn@doi [Publ. Astron. Soc.
  Jap.] {10.1093/pasj/psz080}, 71, Publications of the Astronomical Society of
  Japan, Volume 71, Issue 5, October 2019, 98,
  https://doi.org/10.1093/pasj/psz080

\bibitem[\protect\citeauthoryear{Nozawa, Stergioulas, Gourgoulhon  \&
  Eriguchi}{Nozawa et~al.}{1998}]{Nozawa1998}
Nozawa T.,  Stergioulas N.,  Gourgoulhon E.,   Eriguchi Y.,  1998, Suppl

\bibitem[\protect\citeauthoryear{Okawa, Fujisawa, Yamamoto, Hirai, Yasutake,
  Nagakura  \& Yamada}{Okawa et~al.}{2018}]{Okawa:2018smx}
Okawa H.,  Fujisawa K.,  Yamamoto Y.,  Hirai R.,  Yasutake N.,  Nagakura H.,
  Yamada S.,  2018

\bibitem[\protect\citeauthoryear{Otsuki, Tagoshi, Kajino  \& Wanajo}{Otsuki
  et~al.}{2000}]{Otsuki:1999kb}
Otsuki K.,  Tagoshi H.,  Kajino T.,   Wanajo S.-y.,  2000, \mn@doi [Astrophys.
  J.] {10.1086/308632}, 533, 424

\bibitem[\protect\citeauthoryear{Panov \& Janka}{Panov \&
  Janka}{2009}]{Panov:2008tr}
Panov I.~V.,  Janka H.~T.,  2009, \mn@doi [Astron. Astrophys.]
  {10.1051/0004-6361:200810292}, 494, 829

\bibitem[\protect\citeauthoryear{Paschalidis \& Stergioulas}{Paschalidis \&
  Stergioulas}{2017}]{Paschalidis:2016vmz}
Paschalidis V.,  Stergioulas N.,  2017, \mn@doi [Living Rev. Rel.]
  {10.1007/s41114-017-0008-x}, 20, 7

\bibitem[\protect\citeauthoryear{Pons, Reddy, Prakash, Lattimer  \&
  Miralles}{Pons et~al.}{1999}]{Pons:1998mm}
Pons J.~A.,  Reddy S.,  Prakash M.,  Lattimer J.~M.,   Miralles J.~A.,  1999,
  \mn@doi [Astrophys. J.] {10.1086/306889}, 513, 780

\bibitem[\protect\citeauthoryear{Prakash, Bombaci, Prakash, Ellis, Lattimer  \&
  Knorren}{Prakash et~al.}{1997}]{Prakash:1996xs}
Prakash M.,  Bombaci I.,  Prakash M.,  Ellis P.~J.,  Lattimer J.~M.,   Knorren
  R.,  1997, \mn@doi [Phys. Rept.] {10.1016/S0370-1573(96)00023-3}, 280, 1

\bibitem[\protect\citeauthoryear{Press, Teukolsky, Vetterling  \&
  Flannery}{Press et~al.}{1992}]{Press:1992zz}
Press W.~H.,  Teukolsky S.~A.,  Vetterling W.~T.,   Flannery B.~P.,  1992,
  {Numerical Recipes in FORTRAN: The Art of Scientific Computing}

\bibitem[\protect\citeauthoryear{Rosswog \& Diener}{Rosswog \&
  Diener}{2021}]{Rosswog:2020kwm}
Rosswog S.,  Diener P.,  2021, \mn@doi [Class. Quant. Grav.]
  {10.1088/1361-6382/abee65}, 38, 115002

\bibitem[\protect\citeauthoryear{{Roxburgh}}{{Roxburgh}}{2006}]{Roxburgh2006}
{Roxburgh} I.~W.,  2006, \mn@doi [\aap] {10.1051/0004-6361:20065109}, \href
  {https://ui.adsabs.harvard.edu/abs/2006A&A...454..883R} {454, 883}

\bibitem[\protect\citeauthoryear{{Stergioulas} \& {Friedman}}{{Stergioulas} \&
  {Friedman}}{1995}]{Stergioulas1995}
{Stergioulas} N.,  {Friedman} J.~L.,  1995, \mn@doi [\apj] {10.1086/175605},
  \href {https://ui.adsabs.harvard.edu/abs/1995ApJ...444..306S} {444, 306}

\bibitem[\protect\citeauthoryear{Strobel, Schaab  \& Weigel}{Strobel
  et~al.}{1999}]{Strobel:1999vn}
Strobel K.,  Schaab C.,   Weigel M.~K.,  1999, Astron. Astrophys., 350, 497

\bibitem[\protect\citeauthoryear{{Sumiyoshi}, {Ib{\'a}{\~n}ez}  \&
  {Romero}}{{Sumiyoshi} et~al.}{1999}]{Sumiyoshi1999}
{Sumiyoshi} K.,  {Ib{\'a}{\~n}ez} J.~M.,   {Romero} J.~V.,  1999, \mn@doi
  [\aaps] {10.1051/aas:1999123}, \href
  {https://ui.adsabs.harvard.edu/abs/1999A&AS..134...39S} {134, 39}

\bibitem[\protect\citeauthoryear{Sumiyoshi, Suzuki, Otsuki, Terasawa  \&
  Yamada}{Sumiyoshi et~al.}{2000}]{Sumiyoshi:1999rh}
Sumiyoshi K.,  Suzuki H.,  Otsuki K.,  Terasawa M.,   Yamada S.,  2000, \mn@doi
  [Publ. Astron. Soc. Jap.] {10.1093/pasj/52.4.601}, 52, 601

\bibitem[\protect\citeauthoryear{{Tassoul}}{{Tassoul}}{1978}]{Tassoul:1978}
{Tassoul} J.-L.,  1978, {Theory of rotating stars}

\bibitem[\protect\citeauthoryear{Terasawa, Sumiyoshi, Kajino, Mathews  \&
  Tanihata}{Terasawa et~al.}{2001}]{Terasawa:2001wn}
Terasawa M.,  Sumiyoshi K.,  Kajino T.,  Mathews G.~J.,   Tanihata I.,  2001,
  \mn@doi [Astrophys. J.] {10.1086/323526}, 562, 470

\bibitem[\protect\citeauthoryear{{The Sage Developers}}{{The Sage
  Developers}}{2019}]{sagemath}
{The Sage Developers} 2019, {S}ageMath, the {S}age {M}athematics {S}oftware
  {S}ystem ({V}ersion 8.6.0)

\bibitem[\protect\citeauthoryear{{Uryu} \& {Eriguchi}}{{Uryu} \&
  {Eriguchi}}{1994}]{Uryu:1994}
{Uryu} K.,  {Eriguchi} Y.,  1994, \mn@doi [\mnras] {10.1093/mnras/269.1.24},
  \href {https://ui.adsabs.harvard.edu/abs/1994MNRAS.269...24U} {269, 24}

\bibitem[\protect\citeauthoryear{{Uryu} \& {Eriguchi}}{{Uryu} \&
  {Eriguchi}}{1995}]{Uryu:1995}
{Uryu} K.,  {Eriguchi} Y.,  1995, \mn@doi [\mnras] {10.1093/mnras/277.4.1411},
  \href {https://ui.adsabs.harvard.edu/abs/1995MNRAS.277.1411U} {277, 1411}

\bibitem[\protect\citeauthoryear{Uryu \& Tsokaros}{Uryu \&
  Tsokaros}{2012}]{Uryu:2011ky}
Uryu K.,  Tsokaros A.,  2012, \mn@doi [Phys. Rev. D]
  {10.1103/PhysRevD.85.064014}, 85, 064014

\bibitem[\protect\citeauthoryear{Uryu, Tsokaros, Baiotti, Galeazzi, Taniguchi
  \& Yoshida}{Uryu et~al.}{2017}]{Uryu:2017obi}
Uryu K.,  Tsokaros A.,  Baiotti L.,  Galeazzi F.,  Taniguchi K.,   Yoshida S.,
  2017, \mn@doi [Phys. Rev. D] {10.1103/PhysRevD.96.103011}, 96, 103011

\bibitem[\protect\citeauthoryear{Uryu, Yoshida, Gourgoulhon, Markakis,
  Fujisawa, Tsokaros, Taniguchi  \& Eriguchi}{Uryu et~al.}{2019}]{Uryu:2019ckz}
Uryu K.,  Yoshida S.,  Gourgoulhon E.,  Markakis C.,  Fujisawa K.,  Tsokaros
  A.,  Taniguchi K.,   Eriguchi Y.,  2019, \mn@doi [Phys. Rev. D]
  {10.1103/PhysRevD.100.123019}, 100, 123019

\bibitem[\protect\citeauthoryear{Villain, Pons, Cerda-Duran  \&
  Gourgoulhon}{Villain et~al.}{2004}]{Villain:2003ey}
Villain L.,  Pons J.~A.,  Cerda-Duran P.,   Gourgoulhon E.,  2004, \mn@doi
  [Astron. Astrophys.] {10.1051/0004-6361:20035619}, 418, 283

\bibitem[\protect\citeauthoryear{Wanajo, Kajino, Mathews  \& Otsuki}{Wanajo
  et~al.}{2001}]{Wanajo:2001pu}
Wanajo S.,  Kajino T.,  Mathews G.~J.,   Otsuki K.,  2001, \mn@doi [Astrophys.
  J.] {10.1086/321339}, 554, 578

\bibitem[\protect\citeauthoryear{Witti, Janka, Takahashi  \& Hillebrandt}{Witti
  et~al.}{1992}]{Witti:1992fn}
Witti J.,  Janka H.~T.,  Takahashi K.,   Hillebrandt W.,  1992.

\bibitem[\protect\citeauthoryear{Wongwathanarat, Janka  \&
  Mueller}{Wongwathanarat et~al.}{2013}]{Wongwathanarat:2012zp}
Wongwathanarat A.,  Janka H.~T.,   Mueller E.,  2013, \mn@doi [Astron.
  Astrophys.] {10.1051/0004-6361/201220636}, 552, A126

\bibitem[\protect\citeauthoryear{Woosley, Wilson, Mathews, Hoffman  \&
  Meyer}{Woosley et~al.}{1994}]{Woosley:1994ux}
Woosley S.~E.,  Wilson J.~R.,  Mathews G.~J.,  Hoffman R.~D.,   Meyer B.~S.,
  1994, \mn@doi [Astrophys. J.] {10.1086/174638}, 433, 229

\bibitem[\protect\citeauthoryear{{Yamada} \& {Sato}}{{Yamada} \&
  {Sato}}{1994}]{Yamada:1994}
{Yamada} S.,  {Sato} K.,  1994, \mn@doi [\apj] {10.1086/174724}, \href
  {https://ui.adsabs.harvard.edu/abs/1994ApJ...434..268Y} {434, 268}

\bibitem[\protect\citeauthoryear{Yasutake, {Fujisawa}  \& {Yamada}}{Yasutake
  et~al.}{2015}]{Yasutake:2015}
Yasutake N.,  {Fujisawa} K.,   {Yamada} S.,  2015, \mn@doi [\mnras]
  {10.1093/mnrasl/slu166}, \href
  {https://ui.adsabs.harvard.edu/abs/2015MNRAS.446L..56Y} {446, L56}

\bibitem[\protect\citeauthoryear{Yasutake, Fujisawa  \& Yamada}{Yasutake
  et~al.}{2016}]{Yasutake:2016}
Yasutake N.,  Fujisawa K.,   Yamada S.,  2016, \mn@doi [Mon. Not. Roy. Astron.
  Soc.] {10.1093/mnras/stw2216}, 463, 3705

\bibitem[\protect\citeauthoryear{Zhou, Tsokaros, Rezzolla, Xu  \&
  Ury\={u}}{Zhou et~al.}{2018}]{Zhou:2017xhf}
Zhou E.,  Tsokaros A.,  Rezzolla L.,  Xu R.,   Ury\={u} K.,  2018, \mn@doi
  [Phys. Rev. D] {10.1103/PhysRevD.97.023013}, 97, 023013

\bibitem[\protect\citeauthoryear{Zhou, Kiuchi, Shibata, Tsokaros  \& Uryu}{Zhou
  et~al.}{2021}]{Zhou:2021upu}
Zhou E.,  Kiuchi K.,  Shibata M.,  Tsokaros A.,   Uryu K.,  2021, \mn@doi
  [Phys. Rev. D] {10.1103/PhysRevD.103.123011}, 103, 123011

\makeatother
\end{thebibliography}




\appendix

\onecolumn

\section{Explicit differential forms of Euler and Einstein equations}\label{sec:diffeqs}
In this appendix, we describe the equations that we solve in our formulation.
To obtain the field equations below,
we use SageMath by \cite{sagemath} and SageManifolds packages
by \cite{Gourgoulhon_2015}.
The first order differential equations in the radial and polar angle directions from the Euler equation yield
\begin{eqnarray}
0 = F_{r} &\equiv& \frac{(\Omega-\omega)^2r^2\sin^2\theta-N^2B^2}{P+\veps}\frac{\del P}{\del r}
  -NB^2\frac{\del N}{\del r}
  -\frac{(\Omega-\omega)^2r^2\sin^2\theta}{B}\frac{\del B}{\del r}
  -\frac{(\Omega-\omega)r^2\sin^2\theta}{B}\frac{\del \omega}{\del r}
  +(\Omega-\omega)^2r\sin^2\theta,\label{eq:Euler_r}\\
0 = F_{\theta} &\equiv& \frac{(\Omega-\omega)^2r^2\sin^2\theta-N^2B^2}{P+\veps}\frac{\del P}{\del \theta}
  -NB^2\frac{\del N}{\del \theta}
  -\frac{(\Omega-\omega)^2r^2\sin^2\theta}{B}\frac{\del B}{\del \theta}
  -\frac{(\Omega-\omega)r^2\sin^2\theta}{B}\frac{\del \omega}{\del \theta}
  +(\Omega-\omega)^2r^2\sin\theta\cos\theta.\label{eq:Euler_th}
\end{eqnarray}
As for the Einstein equation, among non-trivial components,
we choose the following four second-order differential equations:
\begin{eqnarray}
 E_{tt} &\equiv& G_{tt} -\frac{8\pi G}{c^4}T_{tt}\nonumber\\
 &=& \frac{A^2\omega^2\sin^2\theta}{NB^2}
  \left(r^2\frac{\del^2N}{\del r^2}+\frac{\del^2N}{\del\theta^2}\right)
  +\left(\frac{N^2A}{r^2}-\frac{A\omega^2\sin^2\theta}{B^2}\right)
  \left(r^2\frac{\del^2A}{\del r^2}+\frac{\del^2A}{\del\theta^2}\right)
  +\frac{N^2A^2}{r^2B^2}
  \left(r^2\frac{\del^2B}{\del r^2}+\frac{\del^2B}{\del\theta^2}\right)\nonumber\\
 && -\frac{A^2\omega\sin^2\theta}{B^2}
  \left(r^2\frac{\del^2\omega}{\del
   r^2}+\frac{\del^2\omega}{\del\theta^2}\right)
  +\frac{A^2\omega\sin^2\theta}{NB^2}
  \left(r^2\frac{\del N}{\del r}\frac{\del \omega}{\del r}
   +\frac{\del N}{\del \theta}\frac{\del \omega}{\del \theta}\right)
  +\frac{3A^2\omega\sin^2\theta}{B^3}
  \left(r^2\frac{\del B}{\del r}\frac{\del \omega}{\del r}
   +\frac{\del B}{\del \theta}\frac{\del \omega}{\del
   \theta}\right)\nonumber\\
  &&-\left(\frac{N^2}{r^2}-\frac{\omega^2\sin^2\theta}{B^2}\right)
  \left\{r^2\left(\frac{\del A}{\del r}\right)^2
   +\left(\frac{\del A}{\del\theta}\right)^2\right\}
-\frac{2N^2A^2}{r^2B^2}
  \left\{r^2\left(\frac{\del B}{\del r}\right)^2
   +\left(\frac{\del B}{\del\theta}\right)^2\right\}
 -\left(\frac{A^2\sin^2\theta}{4B^2}
    +\frac{3r^2\omega^2\sin^4\theta}{4N^2B^4}\right)
  \left\{r^2\left(\frac{\del\omega}{\del r}\right)^2
   +\left(\frac{\del\omega}{\del\theta}\right)^2\right\}\nonumber\\
  &&+\frac{rA^2\omega^2\sin^2\theta}{NB^2}\frac{\del N}{\del r}
  +\left(\frac{N^2A}{r}-\frac{rA\omega^2\sin^2\theta}{B^2}\right)
  \frac{\del A}{\del r} +\frac{2N^2A^2}{rB}\frac{\del B}{\del r}
  -\frac{4rA^2\omega\sin^2\theta}{B^2}\frac{\del\omega}{\del r}
  +\frac{2N^2A^2\cos\theta}{r^2B\sin\theta}\frac{\del B}{\del\theta}
  -\frac{3A^2\omega\sin\theta\cos\theta}{N^2}\frac{\del\omega}{\del\theta}\nonumber\\
 && -\frac{8\pi G}{c^4}
  \left[ \left(\Omega-\omega\right)^2\veps\omega^2r^4\sin^4\theta
   +\left(\Omega^2 P+2\Omega\omega\veps -2\omega^2\veps\right)
   N^2B^2r^2\sin^2\theta +\veps N^4B^4
	    \right],\label{eq:Ett}\\
 E_{\theta\theta} &\equiv& G_{\theta\theta} -\frac{8\pi G}{c^4}T_{\theta\theta}\nonumber\\
 &=& \frac{r^2}{N}\frac{\del^2N}{\del r^2}
  -\frac{r^2}{B}\frac{\del^2B}{\del r^2}
  +\frac{1}{NA}\left(r^2\frac{\del N}{\del r}\frac{\del A}{\del r}
	       +\frac{\del N}{\del \theta}\frac{\del A}{\del \theta}\right)
  -\frac{1}{AB}\left(r^2\frac{\del A}{\del r}\frac{\del B}{\del r}
	       +\frac{\del A}{\del \theta}\frac{\del B}{\del \theta}\right)
  -\frac{1}{BN}
  \left(r^2\frac{\del B}{\del r}\frac{\del N}{\del r}
   +\frac{\del B}{\del \theta}\frac{\del N}{\del \theta}\right)
  +\frac{r}{N}\frac{\del N}{\del r}\nonumber\\
 && +\frac{r}{A}\frac{\del A}{\del r}
  -\frac{2r}{B}\frac{\del B}{\del r}
  +\frac{\cos\theta}{N\sin\theta}\frac{\del N}{\del \theta}
  -\frac{\cos\theta}{A\sin\theta}\frac{\del A}{\del \theta}
  +\frac{2r^2}{B^2}\left(\frac{\del B}{\del r}\right)^2
  -\frac{r^4\sin^2\theta}{4N^2B^2}\left(\frac{\del \omega}{\del r}\right)^2
  +\frac{r^2\sin^2\theta}{4N^2B^2}\left(\frac{\del \omega}{\del \theta}\right)^2
  -\frac{8\pi G}{c^4}r^2 P,\label{eq:Ethth}\\
 E_{\vphi\vphi} &\equiv& G_{\vphi\vphi} -\frac{8\pi G}{c^4}T_{\vphi\vphi}\nonumber\\
 &=& \frac{A^2\sin^2\theta}{NB^2}
  \left(r^2\frac{\del^2N}{\del r^2} +\frac{\del^2N}{\del \theta^2}\right)
  -\frac{A\sin^2\theta}{B^2}
  \left(r^2\frac{\del^2A}{\del r^2} +\frac{\del^2A}{\del\theta^2}\right)
  +\frac{\sin^2\theta}{B^2}
  \left\{r^2\left(\frac{\del A}{\del r}\right)^2 +\left(\frac{\del A}{\del\theta}\right)^2\right\}\nonumber\\
 && -\frac{3A^2r^2\sin^2\theta}{4N^2B^4}
  \left\{r^2\left(\frac{\del \omega}{\del r}\right)^2 +\left(\frac{\del \omega}{\del\theta}\right)^2\right\}
 +\frac{rA^2\sin^2\theta}{NB^2}\frac{\del N}{\del r}
  -\frac{rA\sin^2\theta}{B^2}\frac{\del A}{\del r}
  -\frac{8\pi G}{c^4}\frac{PN^2B^2r^2\sin^2\theta +\veps\left(\Omega-\omega\right)^2r^4\sin^4\theta}{N^2B^4-B^2\left(\Omega-\omega\right)^2r^2\sin^2\theta}
  ,\label{eq:Ephph}\\
 E_{t\vphi} &\equiv& G_{t\vphi} -\frac{8\pi G}{c^4}T_{t\vphi}\nonumber\\
 &=& -\frac{A^2\omega\sin^2\theta}{NB^2}
  \left(r^2\frac{\del^2N}{\del r^2} +\frac{\del^2N}{\del \theta^2}\right)
  +\frac{A\omega\sin^2\theta}{B^2}
  \left(r^2\frac{\del^2A}{\del r^2} +\frac{\del^2A}{\del \theta^2}\right)
  +\frac{A^2\sin^2\theta}{2B^2}
  \left(r^2\frac{\del^2\omega}{\del r^2} +\frac{\del^2\omega}{\del \theta^2}\right)\nonumber\\
 && -\frac{A^2\sin^2\theta}{2NB^2}
  \left(r^2\frac{\del N}{\del r}\frac{\del \omega}{\del r}
   +\frac{\del N}{\del \theta}\frac{\del \omega}{\del \theta}\right)
 -\frac{3A^2\sin^2\theta}{2B^3}
  \left(r^2\frac{\del B}{\del r}\frac{\del \omega}{\del r}
   +\frac{\del B}{\del \theta}\frac{\del \omega}{\del \theta}\right)\nonumber\\
 && -\frac{\omega\sin^2\theta}{B^2}
  \left\{r^2\left(\frac{\del A}{\del r}\right)^2
   +\left(\frac{\del A}{\del \theta}\right)^2\right\}
 +\frac{3A^2\omega r^2\sin^4\theta}{4N^2B^4}
  \left\{r^2\left(\frac{\del \omega}{\del r}\right)^2
   +\left(\frac{\del \omega}{\del \theta}\right)^2\right\}\nonumber\\
 && -\frac{rA^2\omega\sin^2\theta}{NB^2}\frac{\del N}{\del r}
  +\frac{rA\omega\sin^2\theta}{B^2}\frac{\del A}{\del r}
  +\frac{2rA^2\sin^2\theta}{B^2}\frac{\del \omega}{\del r}
  +\frac{3A^2\sin\theta\cos\theta}{2B^2}\frac{\del \omega}{\del \theta}
  \nonumber\\
 && -\frac{8\pi G}{c^4}
  \left[
   \frac{(\Omega-\omega)^2\veps\omega r^4\sin^4\theta
   +\left\{\Omega P+\veps\left(\Omega-\omega\right)\right\}N^2B^2}{\left(\Omega-\omega
   \right)^2B^2r^2\sin^2\theta -N^2B^4}
  \right].\label{eq:Etph}
\end{eqnarray}
Note that the other non-trivial components of the Einstein equation are automatically satisfied through the Bianchi identity.

To solve the outside geometry of rotating stars as well,
we use a compactifed coordinate
\begin{eqnarray}
 s = \frac{r}{r+R_e}
\end{eqnarray}
where $R_e$ is the surface radius in the equator
and $s=1$ corresponds to the spatial infinity
~(\citet{Cook:1992,Stergioulas1995}).
The asymptotic flat condition
is imposed as the boundary condition
for the metric functions at the spatial infinity,
that is,
\begin{eqnarray}
 N=A=B=1, \omega =0.
\end{eqnarray}
Neumann boundary condition is imposed at the origin
~$s=0$,
\begin{eqnarray}
 \frac{\del N}{\del r}=\frac{\del A}{\del r}=\frac{\del B}{\del r}=\frac{\del \omega}{\del r}=0.
\end{eqnarray}

\section{Isoparametric Finite-Element}\label{sec:isoparametric}

\subsection{One-dimensional interpolation by a quadratic function}
\begin{figure}
\begin{tabular}{cc}
 \includegraphics[width=7.6cm,clip]{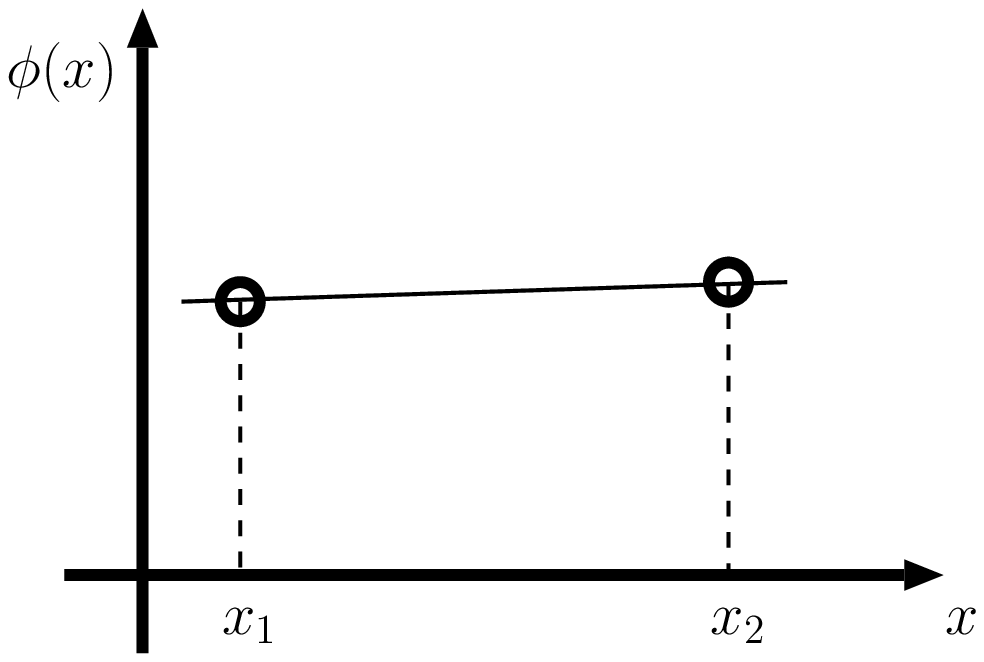}
 &
\includegraphics[width=6.4cm]{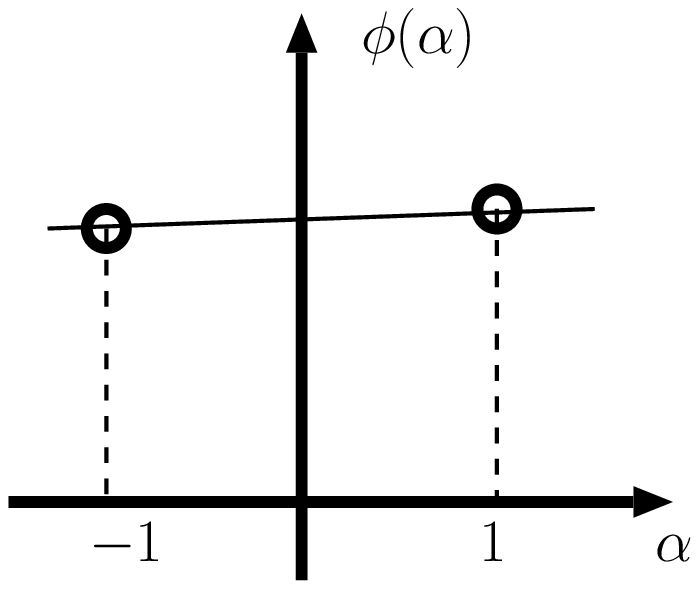}\\
(a) Physical grid & (b) Computational grid
\end{tabular}
\caption{Isoparametric Interpolation using the first-order FE given by a map from (a) a physical coordinate to (b) a computational coordinate.}
\label{fig:iso1st}
\end{figure}
In Fig.~\ref{fig:iso1st}, we define a linear map from
~(a) $x$ coordinate~($x_1\leq x\leq x_2$) 
in the real space and 
~(b) $\alpha$ coordinate~($-1\leq \alpha \leq 1$)
 in the computational space by
\begin{eqnarray}
 x\left(\alpha\right) = a + b\alpha.
\end{eqnarray}
The coefficients~$a$ and $b$ are determined by
the values of two end points 
in the real space:
\begin{eqnarray}
  \left\{
 \begin{array}{lcl}
  x_1 &=& \displaystyle a +b,\\[4mm]
  x_2 &=& \displaystyle a -b,
 \end{array}
 \right.\quad\longrightarrow\quad
 \left\{
 \begin{array}{lcl}
  a &=& \displaystyle \frac{x_1+x_2}{2},\\[4mm]
  b &=& \displaystyle\frac{x_1 -x_2}{2},
 \end{array}
 \right.
\end{eqnarray}
yielding
\begin{eqnarray}
 x &=& \frac{1}{2}\left(1+\alpha\right)x_1
  +\frac{1}{2}\left(1-\alpha\right)x_2
 \equiv \sum_{i=1}^{2} \hat{N}_{i}\left(\alpha\right) x_i,
\label{eq:iso1st_x}
\end{eqnarray}
where the so-called shape functions are given
\begin{eqnarray}
 \hat{N}_1\left(\alpha\right) = \frac{1}{2}\left(1+\alpha\right),\quad
  \hat{N}_2\left(\alpha\right) = \frac{1}{2}\left(1-\alpha\right).
\end{eqnarray}

Any function~$\phi\left(x\right)$ 
can be expanded by the same shape functions
 as
\begin{eqnarray}
 \phi\left(x\right) &=& \sum_{i=1}^{2} \hat{N}_i\left(\alpha\right) \phi_i,
\label{eq:iso1st_f}
\end{eqnarray}
where $\phi_i$ are values at $x=x_i$.
The derivatives with respect to the coordinate~$x$
are evaluated in terms of the derivatives with respect to
 $\alpha$ by
\begin{eqnarray}
 \frac{\dif\phi}{\dif x} &=& \frac{\dif\alpha}{\dif x}
  \sum_{i=1}^{2}\frac{\dif \hN_i}{\dif\alpha}\phi_i
  = \left(\frac{\dif x}{\dif\alpha}\right)^{-1}
  \sum_{i=1}^{2}\frac{\dif \hN_i}{\dif\alpha}\phi_i,
\end{eqnarray}
where
\begin{eqnarray}
 \left(\frac{\dif x}{\dif\alpha}\right)^{-1} =
  \frac{1}{\sum_{i=1}^{2}\frac{\dif \hN_i}{\dif\alpha}x_i}.
\end{eqnarray}
Therefore the first derivative in the finite-element~(FE) is
\begin{eqnarray}
 \frac{\dif\phi}{\dif x} = \frac{\phi_2-\phi_1}{x_2-x_1},
\end{eqnarray}
which exactly coincides with the derivative of the first-order finite-difference.

The function~$\phi$
is integrated in the computational grid as follows.
\begin{eqnarray}
 \int_{x_1}^{x_2} \phi \dif x
  =\int_{-1}^{1} \phi(x(\alpha)) \frac{\dif x}{\dif \alpha}\dif \alpha
= \sum_{i,j}^{2} x_i\phi_j \int_{-1}^{1}\hat{N}_j
\frac{\dif\hat{N}_i}{\dif \alpha}\dif\alpha
= \sum_{i,j}^{2} x_i\phi_j \frac{(-1)^i}{2}
\int_{-1}^{1}\hat{N}_j\dif \alpha
= \sum_{i,j}^{2} x_i\phi_j \frac{(-1)^i}{2}
= \frac{\phi_1+\phi_2}{2}\left(x_2-x_1\right),
\end{eqnarray}
which is the same result as the integration by the trapezoidal rule.

\subsection{One-dimensional interpolation with a cubic function}
\begin{figure}
\begin{tabular}{cc}
 \includegraphics[width=7.6cm,clip]{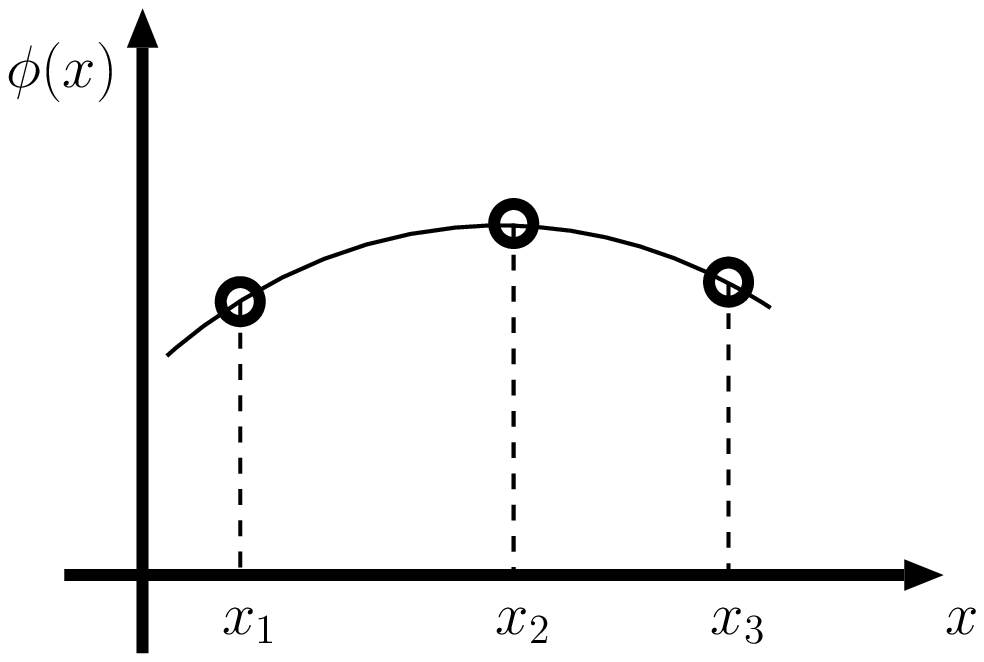}
 &
\includegraphics[width=5.8cm]{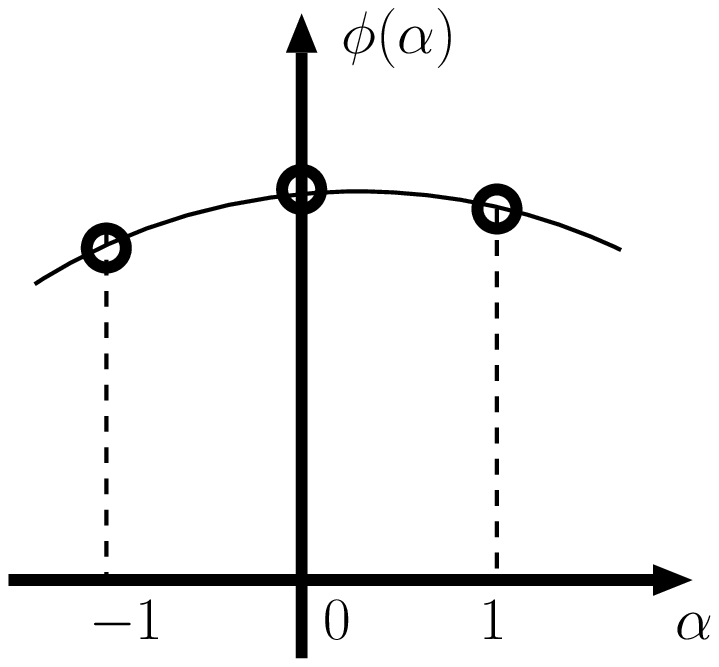}\\
(a) Physical grid & (b) Computational grid
\end{tabular}
\caption{Isoparametric Interpolation using the second-order finite-element given by a map from (a) a physical coordinate to (b) a computational coordinate.}
\label{fig:iso2nd}
\end{figure}
In Fig.~\ref{fig:iso2nd}, we define a higher-order map from~(a) $x$ coordinate~($x_1\leq x\leq x_3$)
in the real space and 
~(b) $\alpha$ coordinate~($-1\leq \alpha \leq 1$)
 in the computational space by
\begin{eqnarray}
 x\left(\alpha\right) = a + b\alpha +c\alpha^2.
\end{eqnarray}
The coefficients~($a, b, c$) are determined by taking three points
in the real space~($x_1, x_2, x_3$):
\begin{eqnarray}
  \left\{
 \begin{array}{lcl}
  x_1 &=& \displaystyle a -b +c,\\[4mm]
  x_2 &=& \displaystyle a,\\[4mm]
  x_3 &=& \displaystyle a +b +c,
 \end{array}
 \right.\quad\longrightarrow\quad
 \left\{
 \begin{array}{lcl}
  a &=& \displaystyle x_2,\\[4mm]
  b &=& \displaystyle \frac{x_3-x_1}{2},\\[4mm]
  c &=& \displaystyle\frac{x_3 -2x_2 +x_1}{2},
 \end{array}
 \right.
\end{eqnarray}
yielding
\begin{eqnarray}
 x &=& -\frac{\alpha}{2}\left(1-\alpha\right)x_1
  +\left(1+\alpha\right)\left(1-\alpha\right)x_2
  +\frac{\alpha}{2}\left(1+\alpha\right)x_3
 \equiv \sum_{i=1}^{3} \hat{N}_{i}\left(\alpha\right) x_i,
\label{eq:iso2nd_x}
\end{eqnarray}
where the shape functions are given
\begin{eqnarray}
 \left\{
 \begin{array}{lcl}
  \hat{N}_1\left(\alpha\right) &=& \displaystyle -\frac{\alpha}{2}\left(1-\alpha\right),\\[4mm]
  \hat{N}_2\left(\alpha\right) &=& \displaystyle \left(1+\alpha\right)\left(1-\alpha\right),\\[4mm]
  \hat{N}_3\left(\alpha\right) &=& \displaystyle \frac{\alpha}{2}\left(1+\alpha\right).
 \end{array}
 \right.
\end{eqnarray}
Any function~$\phi$ is expanded using the shape functions
as bases:
\begin{eqnarray}
 \phi\left(x\right) &=& \sum_{i=1}^{3} \hat{N}_i\left(\alpha\right) \phi_i,
\label{eq:iso2nd_f}
\end{eqnarray}
where $\phi_i$ are values at $x=x_i$.
The first and second spatial derivatives 
are calculated by
\begin{eqnarray}
 \frac{\dif\phi}{\dif x} &=& \frac{\dif\alpha}{\dif x}
  \sum_{i=1}^{3}\frac{\dif \hN_i}{\dif\alpha}\phi_i
  = \left(\frac{\dif x}{\dif\alpha}\right)^{-1}
  \sum_{i=1}^{3}\frac{\dif \hN_i}{\dif\alpha}\phi_i,\\
  \frac{\dif^2\phi}{\dif x^2} &=&
   -\left(\frac{\dif x}{\dif\alpha}\right)^{-3}\frac{\dif^2 x}{\dif\alpha^2}
   \sum_{i=1}^{3}\frac{\dif \hN_i}{\dif\alpha}\phi_i
   +\left(\frac{\dif x}{\dif\alpha}\right)^{-2}
   \sum_{i=1}^{3}\frac{\dif^2 \hN_i}{\dif\alpha^2}\phi_i,
\end{eqnarray}
where
the derivatives of the shape functions are
\begin{eqnarray}
 \left\{
 \begin{array}{lcl}
  \displaystyle \frac{\del\hN_1}{\del \alpha}\left(\alpha\right) &=&
   \displaystyle -\frac{1}{2} +\alpha,\\[4mm]
  \displaystyle \frac{\del\hN_2}{\del \alpha}\left(\alpha\right) &=&
  \displaystyle -2\alpha,\\[4mm]
  \displaystyle \frac{\del\hN_3}{\del \alpha}\left(\alpha\right) &=&
  \displaystyle \frac{1}{2} +\alpha,
 \end{array}
 \right.\hspace{2cm}
 \left\{
 \begin{array}{lcl}
  \displaystyle \frac{\del^2\hN_1}{\del \alpha^2}\left(\alpha\right) &=& \displaystyle 1,\\[4mm]
  \displaystyle \frac{\del^2\hN_2}{\del \alpha^2}\left(\alpha\right) &=& \displaystyle -2,\\[4mm]
  \displaystyle \frac{\del^2\hN_3}{\del \alpha^2}\left(\alpha\right) &=& \displaystyle 1.
 \end{array}
 \right.
\end{eqnarray}

\section{W4IX Method}\label{app:W4IX}
We incorporate finite-element derivatives into the Einstein equation
~\eqref{eq:Ett}-\eqref{eq:Etph} and obtain the system of nonlinear equations.
can be 
It is usually solved by a nonlinear root-finding scheme
 like the Newton-Raphson(NR) method,
which guarantees to find the solution when an initial guess
sufficiently close to the solution is provided.
We proposed a new root-finding scheme, the W4 method,
to greatly improve the initial guess 
dependence~(\citet{Okawa:2018smx,Fujisawa:2018dnh}),
We briefly describe below the W4 method used for solving the metric functions.
A root-finder solves $N$ variables~($x_i$) given $N$ equations($F_i$),
\begin{eqnarray}
 F_i(x_1, x_2, \cdots, x_N) = 0,
\end{eqnarray}
where $i$ runs from $1$ to $N$.
Iterative methods such as the NR method find a better approximate
solution from the current approximate solution.
The damped Newton or line-search method
is a simple extension of the NR method:
\begin{eqnarray}
 x_j^{n+1} &=& x_j^{n} -\Delta\tau\sum_{i=1}^{N}J^{-1}_{ij}F_{i},
\end{eqnarray}
where $0<\Delta\tau<1$, the superscript~$n$ denotes the iterative step
and $J^{-1}_{ij}$ is the inverse of the Jacobian matrix whose component is defined by
\begin{eqnarray}
 J_{ij} &=& \frac{\del F_{i}}{\del x_{j}}.
\end{eqnarray}
Note that it reverts to the NR method when $\Delta\tau=1$.
For visibility, 
we express the damped Newton method as the differential equation
in the matrix form replacing $(x_{n+1}-x_n)/\Delta\tau$ with
$\dot x$ by
\begin{eqnarray}
 \dot{\bx} = -J^{-1}\bF.
\end{eqnarray}
By contrast, in the W4 method, we introduces the second derivative
with respect to the time coordinate~$\tau$ as done
 for the Poisson solver in~\cite{Hirai:2016ogc}
and solve instead the following system:
\begin{eqnarray}
 \dot{\bx} = X\bp,\quad
 \dot{\bp} = -2\bp -Y\bF,\label{eq:W4evolution}
\end{eqnarray}
where $\bp$ is an auxiliary variable and $X,Y$ are the matrices related to the Jacobian matrix.
In the W4 method with the UL decomposition that divides the Jacobian matrix
into an upper matrix and a lower matrix~$J=UL$, we assign them to the W4 matrices as~$X=L^{-1}$ and $Y=U^{-1}$.
The factorization needs $\mathcal{O}(N^3)$ calculations
as the original NR method does.
In what follows, we explain the outline of a new method, W4IX method, to safely reduce the number of calculations.
An important fact is that the Jacobian matrix for our system of nonlinear equations
is a sparse matrix.
With this situation,
we employ the incomplete UL factorization in the same procedure as that of the incomplete LU factorization~(e.g.~\cite{Press:1992zz}).
The incomplete LU or UL decomposition helps to reduce the condition number of the Jacobian matrix in the linear system.
Usually, the system of linear equations including an ill-conditioned matrix, ~$A\bx=\bb$, is transformed into $MA\bx=M\bb$ by
multiplying a pre-conditioning matrix~$M$
whose condition number is smaller than that of the original one.
Since the cost to approximately factorize the sparse matrix is $\mathcal{O}(N^2)$ calculations only,
we approximately decompose the Jacobian matrix into $\hat U$ and $\hat L$ matirices,
which assumes the same pattern of non-zero components 
in the lower and upper matrices as that in the original Jacobian matrix.
The difference between the original Jacobian matrix
and the product of matrices~$\hat U$ and $\hat L$
is defined by
\begin{eqnarray}
 R &=& \hat{U}\hat{L} -J.
\end{eqnarray}
We adopt one of the W4 matrices 
by~$Y\equiv\hat{L}^{-1}\hat{U}^{-1}$.
To compute $Y\bF=\hat{L}^{-1}\hat{U}^{-1}\bF$,
we also pay $\mathcal{O}(N^2)$ calculation cost,
because $\bz=\hat{L}^{-1}\by$ and $\by=\hat{U}^{-1}\bF$
are solved by the forward and backward substitutions, respectively.
The W4 method guarantees the local convergence
when the other W4 matrix~$X$ satisfies the condition~$YJX=E$
~(\cite{Okawa:2018smx}):
\begin{eqnarray}
 X = \left[YJ\right]^{-1} = \left[\hat{L}^{-1}\hat{U}^{-1}J\right]^{-1}
  = \left[E-\hat{L}^{-1}\hat{U}^{-1}R\right]^{-1}.
\end{eqnarray}
If the product of matrices $\hat U$ and $\hat L$ sufficiently reproduces the original Jacobian matrix,
the part~$\hat{L}^{-1}\hat{U}^{-1}R$ is small compared to the identity matrix~$E$ and
thus the W4 matrix $X$ is expressed by
the geometric series 
\begin{eqnarray}
  X = \left[E-\hat{L}^{-1}\hat{U}^{-1}R\right]^{-1}
 = E +\sum_{m=1}^{\infty}\left[\hat{L}^{-1}\hat{U}^{-1}R\right]^m.
\end{eqnarray}
Our concern is the numerical cost of the W4 iteration.
The right hand side of the evolution equation for $\bx$
in Eq.~\eqref{eq:W4evolution} yields
\begin{eqnarray}
 X\bp
  &=& \bp +\sum_{m=1}^{M}\left[\hat{L}^{-1}\hat{U}^{-1}R\right]^m\bp
  = \bp +\sum_{m=1}^{M}\left[E-\hat{L}^{-1}\hat{U}^{-1}J\right]^m\bp,
\end{eqnarray}
where we truncate the series at a finite number~$M$.
Therefore, the calculation cost is $\mathcal{O}(MN^2)$
for the matrix-vector product.
Finally, we summarize the W4 evolution equation
for solving the metric components in the stationary Einstein equation:
\begin{eqnarray}
 \dot{\bx} &=& \bp +\sum_{m=1}^{M}\left[\hat{L}^{-1}\hat{U}^{-1}R\right]^m\bp,\\
 \dot{\bp} &=& -2\bp -\hat{L}^{-1}\hat{U}^{-1}\bF.
\end{eqnarray}

\begin{figure}
 \begin{tabular}{cc}
  \begin{minipage}{0.42\textwidth}
  \includegraphics[width=8.cm,clip]{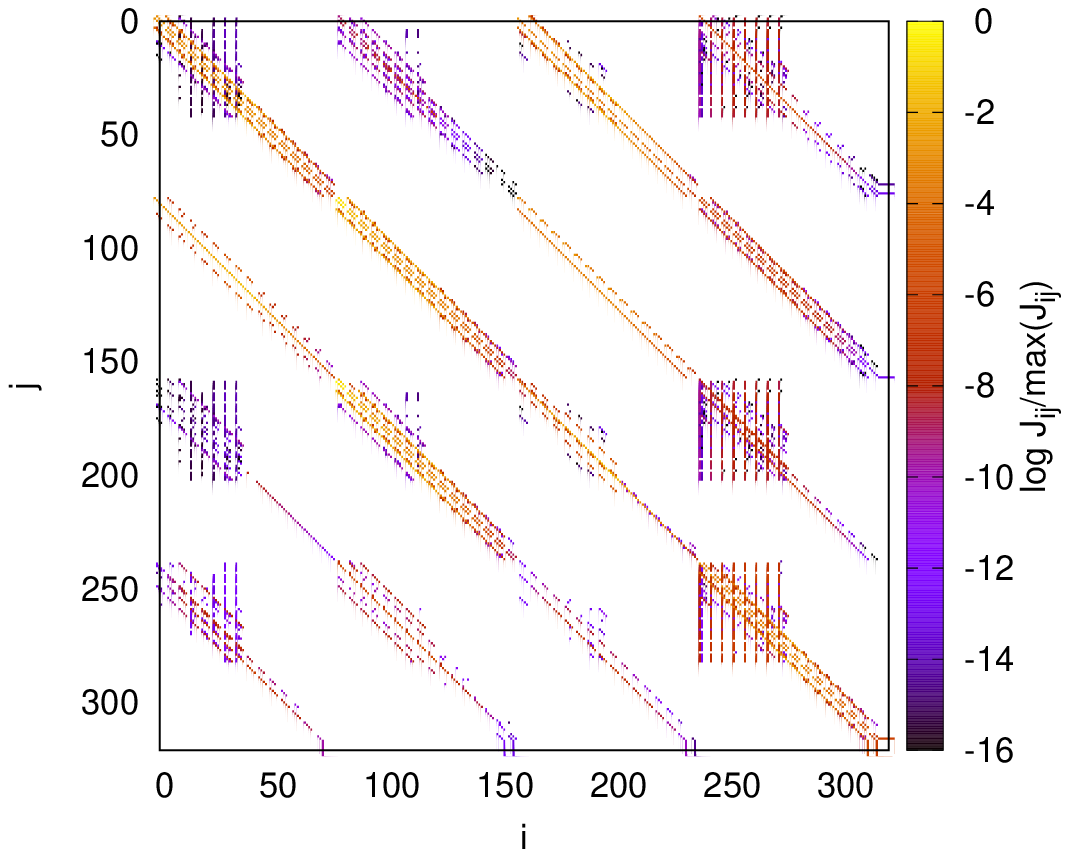}
  \end{minipage} &
  \begin{minipage}{0.6\textwidth}
   \begin{tabular}{cc}
    \includegraphics[width=4.8cm,clip]{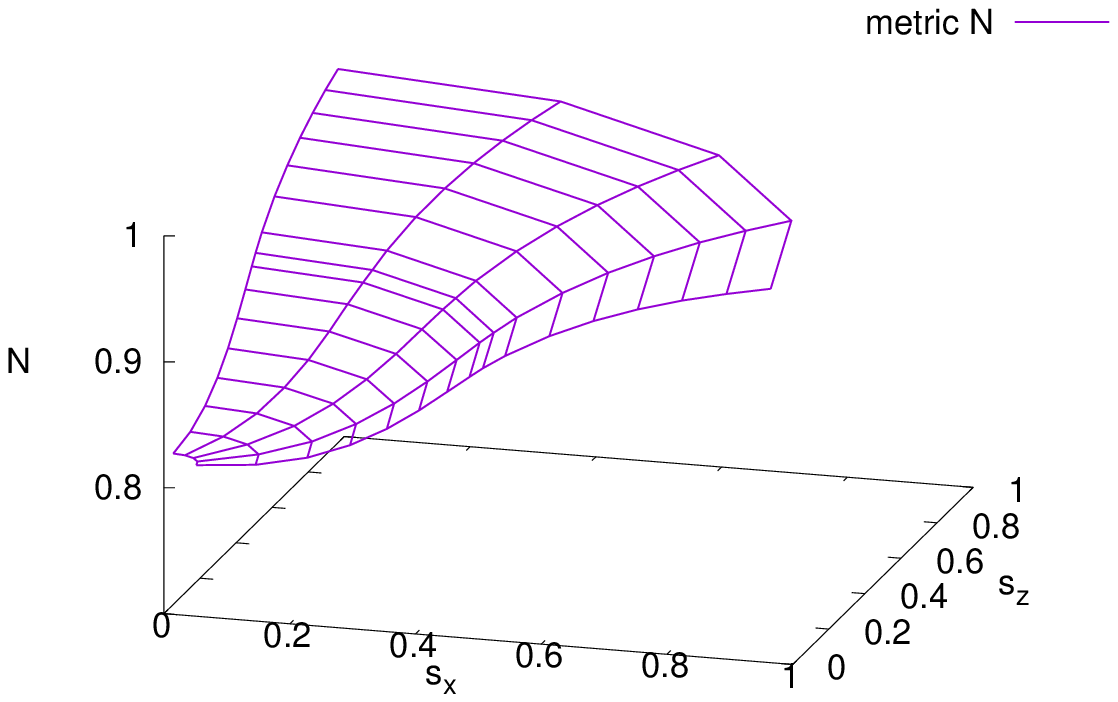} &
	\includegraphics[width=4.8cm,clip]{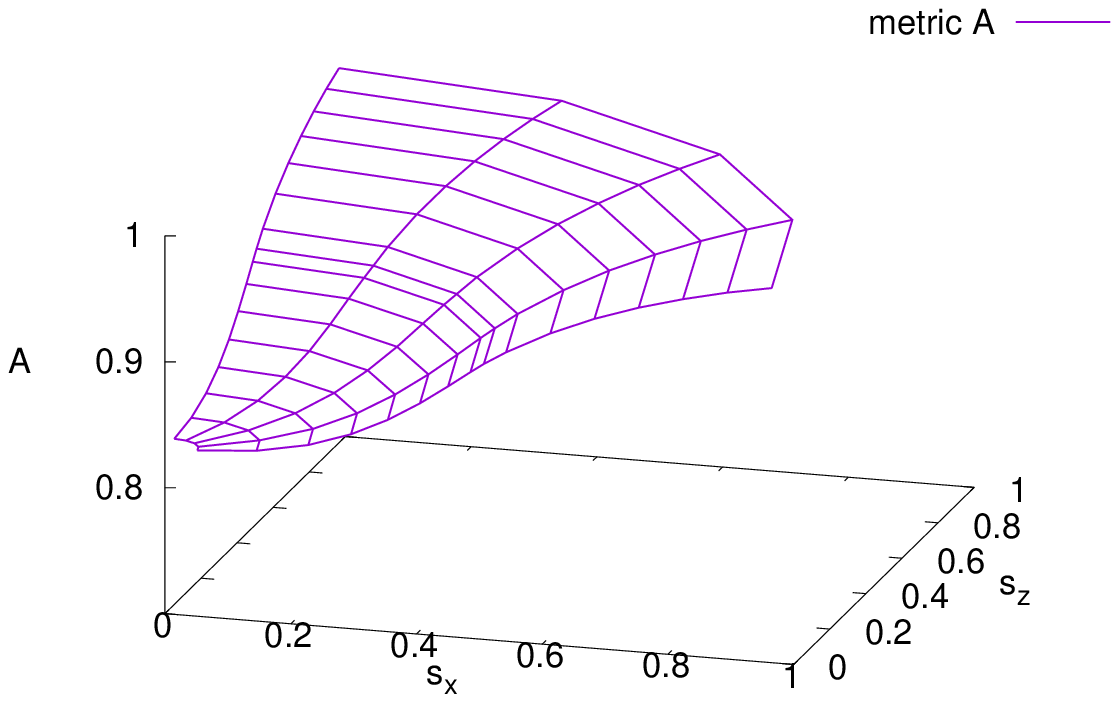} \\
    \includegraphics[width=4.8cm,clip]{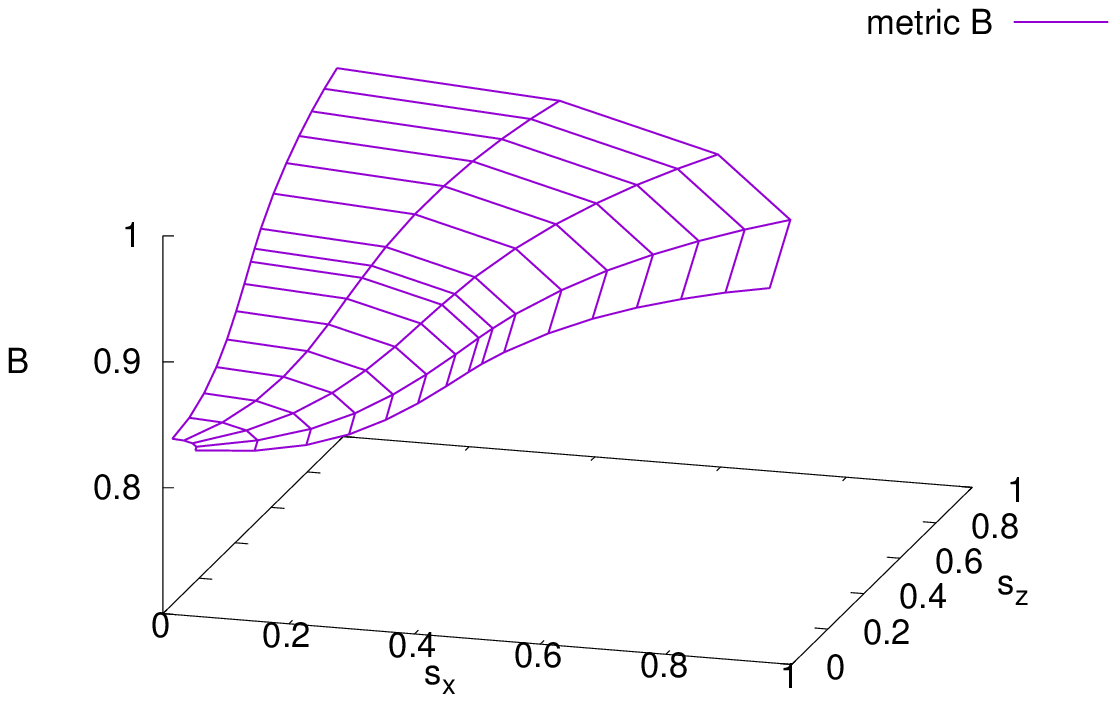} &
	\includegraphics[width=4.8cm,clip]{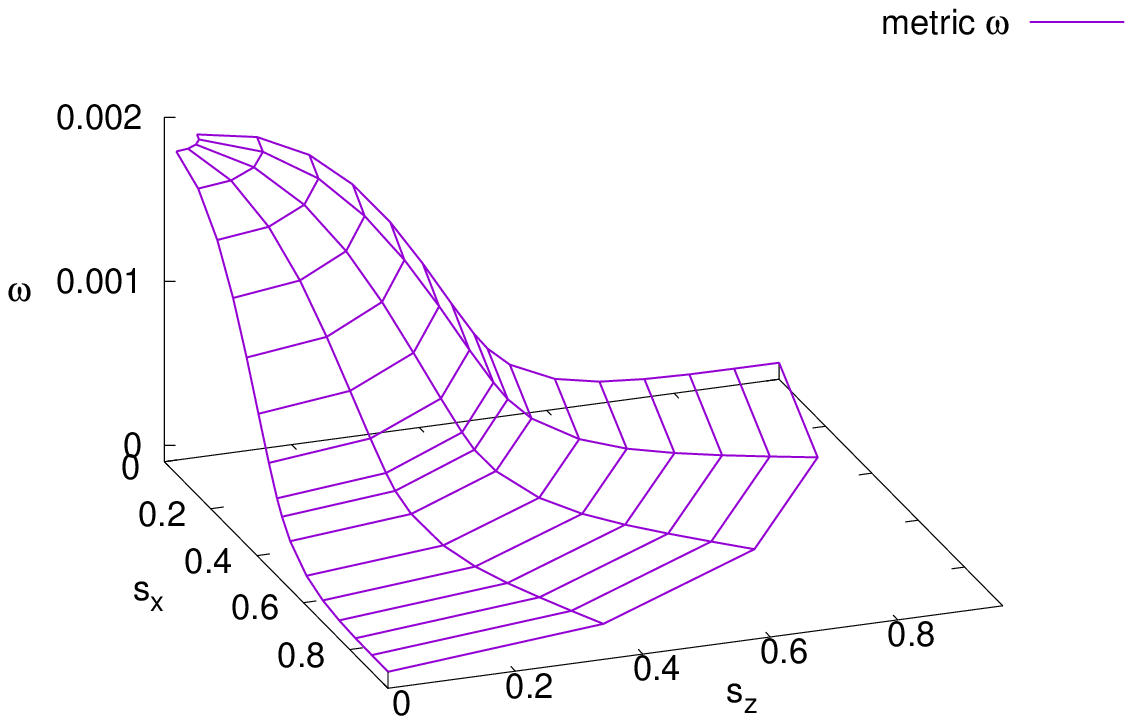}
   \end{tabular} 
  \end{minipage}\\
  (a) & (b)
 \end{tabular} 
 \caption{(a) Non-zero components of Jacobian matrix
 normalized by the maximum value of the components.
 (b) An example of metric functions obtained 
 by solving the discretized the Einstein equation.
 }
 \label{fig:Jacobian_metric}
\end{figure}

As a concrete application of the W4IX method to the metric functions
 of rotating stars described in App.~\ref{sec:diffeqs},
 we consider the low-resolution model corresponding to 
the stellar mesh~$N_r\times N_\theta=8\times 5$.
The outside of stars is also divided by the same number of meshes
as that of the inside.
Schematically, the associated Jacobian matrix
is described by $16$ matrices as
\begin{eqnarray}
J=
 \begin{pmatrix}
  \frac{\del\mathbf{E}_{tt}}{\del\mathbf{N}} &
  \frac{\del\mathbf{E}_{tt}}{\del\mathbf{A}} &
  \frac{\del\mathbf{E}_{tt}}{\del\mathbf{B}} &
  \frac{\del\mathbf{E}_{tt}}{\del\mathbf{\omega}}\\[3mm]
  \frac{\del\mathbf{E}_{\theta\theta}}{\del\mathbf{N}} &
  \frac{\del\mathbf{E}_{\theta\theta}}{\del\mathbf{A}} &
  \frac{\del\mathbf{E}_{\theta\theta}}{\del\mathbf{B}} &
  \frac{\del\mathbf{E}_{\theta\theta}}{\del\mathbf{\omega}}\\[3mm]
  \frac{\del\mathbf{E}_{\vphi\vphi}}{\del\mathbf{N}} &
  \frac{\del\mathbf{E}_{\vphi\vphi}}{\del\mathbf{A}} &
  \frac{\del\mathbf{E}_{\vphi\vphi}}{\del\mathbf{B}} &
  \frac{\del\mathbf{E}_{\vphi\vphi}}{\del\mathbf{\omega}}\\[3mm]
  \frac{\del\mathbf{E}_{t\vphi}}{\del\mathbf{N}} &
  \frac{\del\mathbf{E}_{t\vphi}}{\del\mathbf{A}} &
  \frac{\del\mathbf{E}_{t\vphi}}{\del\mathbf{B}} &
  \frac{\del\mathbf{E}_{t\vphi}}{\del\mathbf{\omega}}
 \end{pmatrix},
\end{eqnarray}
where each block is a $80\times 80$ matrix.
Fig.~\ref{fig:Jacobian_metric}~(a) shows
the non-zero components of the typical Jacobian matrix computed numerically.
They are normalized by the maximum value among them.
While the Jacobian matrix is sparse, 
non-diagonal components are not small,
which makes the problem difficult.
In fact, we confirm that 
the standard NR method does not work
when the initial condition is slightly far from the solution
but the W4 method including the new W4IX method
works well.
In Fig.~\ref{fig:Jacobian_metric}~(b),
we exemplify the metric functions of a slowly rotating star,
which the boundary conditions are actually satisfied
at $s_x=s_z=1$ corresponding to $r=\infty$.

\section{Slice-Shooting Scheme}\label{app:sliceshooting}
\begin{figure}
\begin{tabular}{cc}
\includegraphics[width=7.5cm,clip]{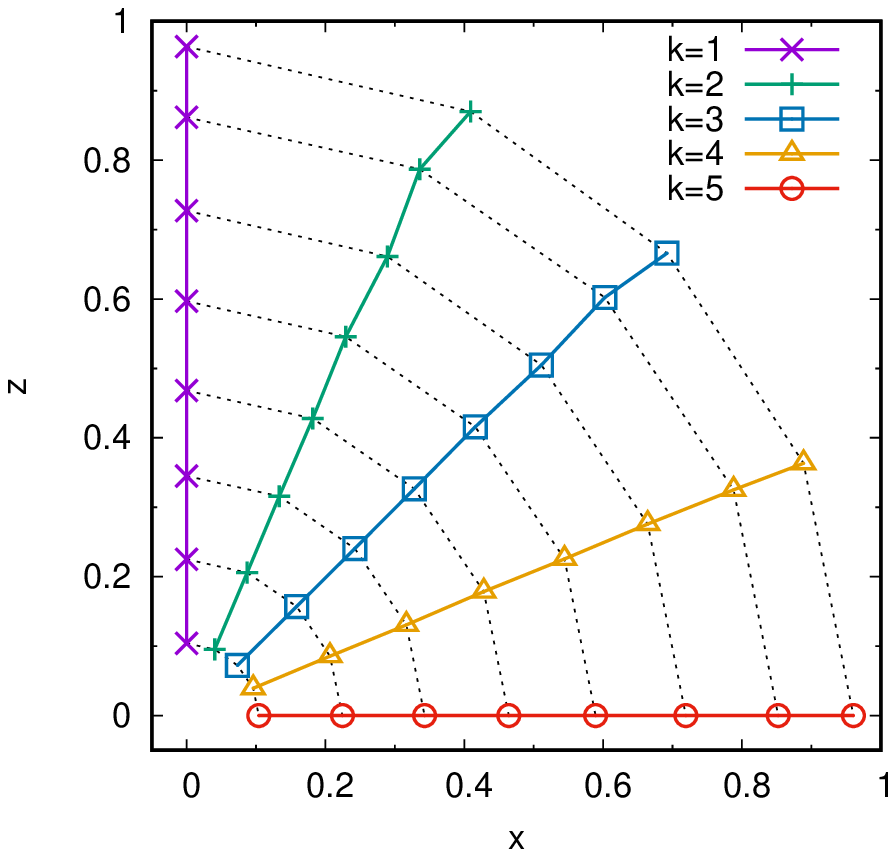}
 &
\includegraphics[width=7.5cm,clip]{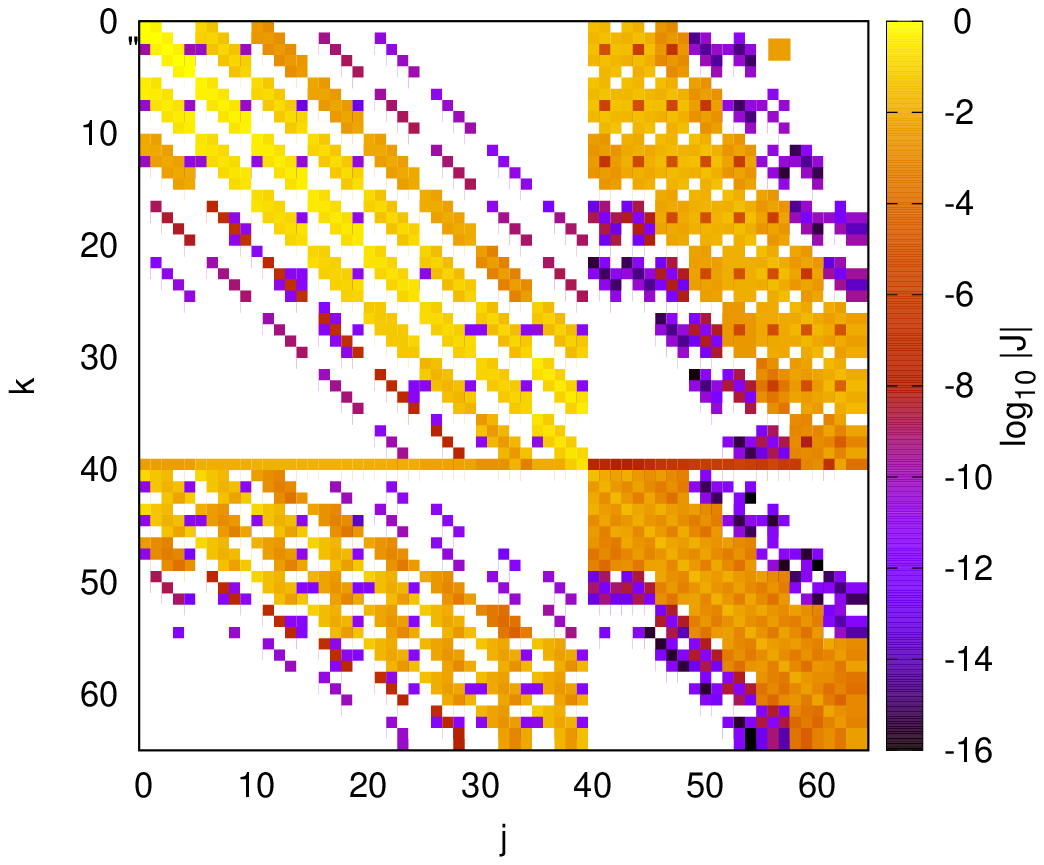}\\
(a) & (b)
\end{tabular} 
 \caption{(a) Schematic picture of the Slice-Shooting method
for the $N_r\times N_{\theta}=8\times 5$ case.
 The same mark points in a slice are solved at the same time with other points fixed.  The slice will be solved from the axis($k=1$) to the equator($k=5$) and again from the equator($k=5$) to the axis($k=1$).  The iteration will be repeated until the convergence is achived.  (b) Jacobian matrix associated with the discretized form of the relativistic Euler equation.  Non-diagonal components are not small and it makes the condition-number higher.
 }
 \label{fig:sliceshooting_def}
\end{figure}
In this section, we describe how to solve the Euler equation in the differential form.
As mentioned in Sec.~\ref{sec:method}, 
we solve the polar coordinates~$r$ and $\theta$ themselves,
which is displayed in Fig.~\ref{fig:sliceshooting_def}~(a)
for the low-resolution model~$N_r\times N_\theta=8\times 5$.
the variables are the coordinates shown with marks.
Since the angles at the rotational axis denoted by $k=1$ and at the equator denoted by $k=5$
can be explicitly given by $\theta=0$ and $\theta=\pi/2$ respectively,
the number of variables for the radial coordinate~$r$ is $8\times 5=40$
while that for the polar angle coordinate~$\theta$ is $8\times (5-2)=24$.
The total number depending on the resolution is shown
in Table~\ref{table:cond}.
In contrast to the Einstein equation,
the Jacobian matrix associated with the discretized Euler equation
by the finite-element is rather dense as Fig.~\ref{fig:sliceshooting_def}~(b):
\begin{eqnarray}
J=
 \begin{pmatrix}
  \frac{\del\mathbf{F}_{r}}{\del\mathbf{r}} &
  \frac{\del\mathbf{F}_{r}}{\del\mathbf{\theta}}\\[3mm]
  \frac{\del\mathbf{F}_{\theta}}{\del\mathbf{r}} &
  \frac{\del\mathbf{F}_{\theta}}{\del\mathbf{\theta}}
 \end{pmatrix}.
\end{eqnarray}
In Table~\ref{table:cond},
we show the condition-number of the Jacobian matrix with each
resolution\footnote{The condition-number depends on the initial condition of iteration schemes.  In this case, we evaluate it when solving a differentially rotating solution from the initial condition which is given by the spherically symmetric solution.},
 which is defined by the ratio of the maximum
 to minimum eigenvalues of the matrix.
A huge condition-number often causes
either the extreme slowdown or the failure
 of iterative schemes.

%

Reducing the size of the Jacobian matrix generally
helps to reduce not only the condition-number of the matrix
but also the computational cost of iterative
 schemes\footnote{This does not guarantee that the total computational cost is reduced, because another iteration is needed until the whole convergence is achieved.}.
The whole variables, the polar coordinates~$r$ and $\theta$, 
are divided into several subgroups.
We explain our new scheme, the slice-shooting scheme,
by th low-resolution model~$N_r\times N_\theta = 8\times 5$.
In Fig.~\ref{fig:sliceshooting_def}~(a), 
we divide the whole into the $5$ slices along the radial direction denoted by the same marks and solve the variables in each slice
with those in the other slices fixed.
In other words, we first solve only the radial coordinates in the slice denoted by the crosses~($k=1$) keeping the other coordinates fixed.
In this particular turn, we solve $8$ variables of the radial coordinates~$r_{j1}$ at the rotational axis by the $8$ radial equations of the Euler equation, which gives the $8\times 8$ Jacobian matrix.
Next, we consider the slice shown by the pluses~($k=2$)
to solve both coordinates~$r_{j2}$ and $\theta_{j2}$ by
the $8$ radial and $8$ angle equations
leading to the $16\times 16$ Jacobian matrix.
The radial and polar angle coordinates at the square points~($k=3$) follow them at the plus points~($k=2$) and so forth.
As such, the original $64\times 64$ Jacobian matrix
is reduced to $8\times 8, 16\times 16, 16\times 16, 16\times 16$ and $8\times 8$ Jacobian matrices.
After solving the radial coordinates~$r_{j5}$ in the slice in the equator with the circles~($k=5$),
we again solve the variables in the slice from the triangles~($k=4$), 
to the crosses~($k=1$), which corresponds to one cycle in the slice-shooting scheme.
Until the whole convergence is achived,
the iteration will continue.

\begin{figure}
\begin{tabular}{cc}
\includegraphics[width=7.5cm,clip]{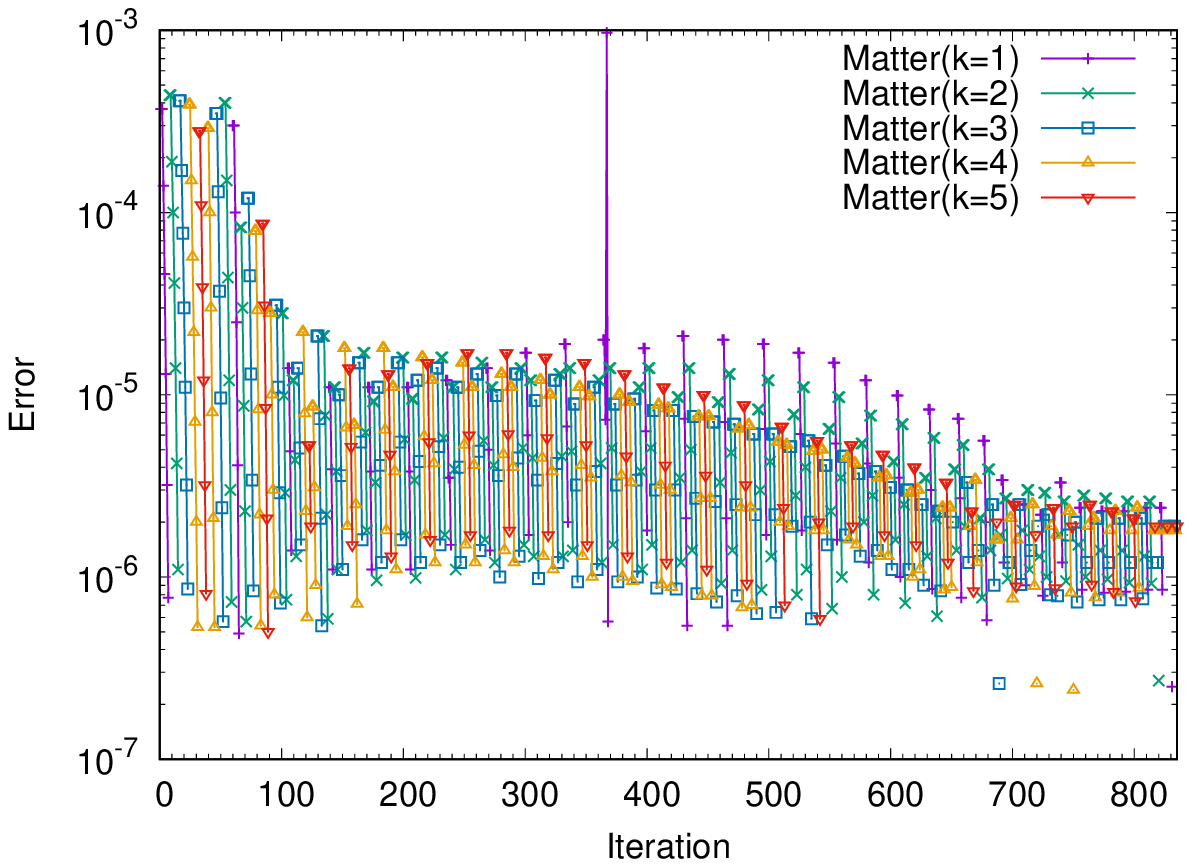}
 &
\includegraphics[width=7.5cm,clip]{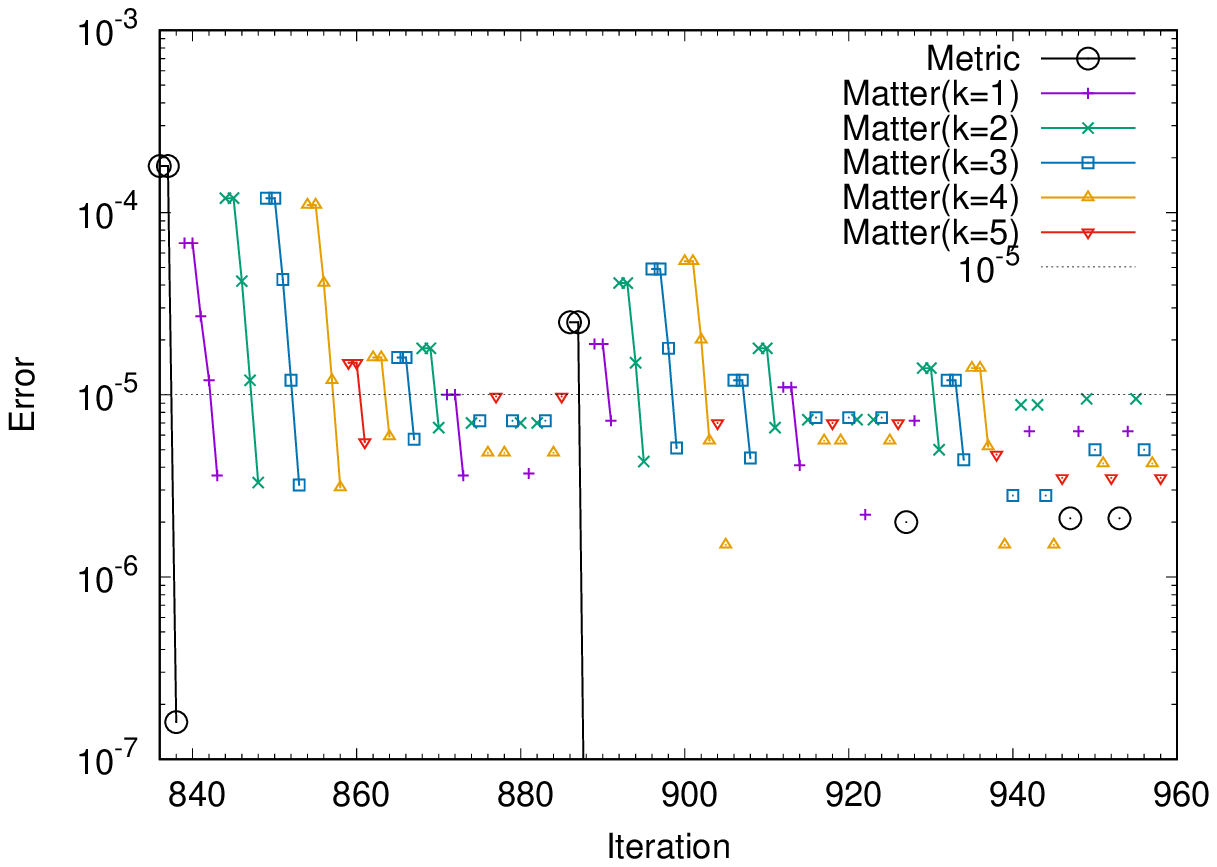}\\
(a) & (b)
\end{tabular} 
 \caption{Error in the Euler equation
 as a function of iteration step.
 Each mark corresponds to the mark in Fig.~\ref{fig:sliceshooting_def}.
 (a) All the errors in the slices with the metric fixed become small after the iteration.
 (b) While the metric functions are solved under the newly obtained coordinates, it does not significantly affect the error in the Euler equation,
which results in decreasing all the errors in the Einstein and Euler equations.
 }
 \label{fig:sliceshooting}
\end{figure}

In Fig.~\ref{fig:sliceshooting}~(a), we show the maximum error in the Euler equation evaluated at each iteration step
starting from the initial guess of the spherically symmetric solution.
Three phases are observed globally:
(i) Rapidly decreasing phase up to around $100$ iteration steps,
(ii) Slowly decreasing phase up to around $700$ iteration steps,
(iii) Converged phase up to $\sim 830$ iteration steps.
In the phase~(ii), 
the error gradually decreases except for the burst-like increase at around $360$ iterations.
In Fig.~\ref{fig:sliceshooting}~(b),
we show not only the maximum error in the Euler equation
but also that in the Einstein equation shown by the black circles.
Once all iterations in the matter section converge in Fig.~\ref{fig:sliceshooting}~(a),
the iteration in the metric section also starts with the first black circle in Fig.~\ref{fig:sliceshooting}~(b).
Note that the convergence in each slice is achieved at most by six iteration steps, while the whole convergence needs around $800$ steps in the matter section.
Fig.~\ref{fig:sliceshooting}~(b) shows the errors 
both in the Einstein and Euler equations successfully decrease
by the slice-shooting scheme.

%
\begin{table}
\centering
\begin{tabular}{ccccc}
 Model & $N_r$ & $N_{\theta}$ & Total Number of Variables & Condition Number $\rho_J$ \\\hline
 $4\times 3$ & $12$ & $4$ & $16$ & $\sim 10^2$ \\\hline
 $8\times 5$ & $40$ & $24$ & $64$ & $\sim 10^5$ \\\hline
 $16\times 9$ & $144$ & $112$ & $256$ & $\sim 10^{7}$ \\\hline
\end{tabular}
\caption{Dependence of the condition number on the size of problem.  The model indicates the number of grid points and the total number of variables for the W4 iteration is given by the numbers of the radial position~$r_{jk}$ and the polar angle~$\theta_{jk}$.}
 \label{table:cond}
\end{table}

\section{Rezoning}\label{sec:rezoning}
\begin{figure}
 \begin{tabular}{cc}
  \includegraphics[width=7.5cm]{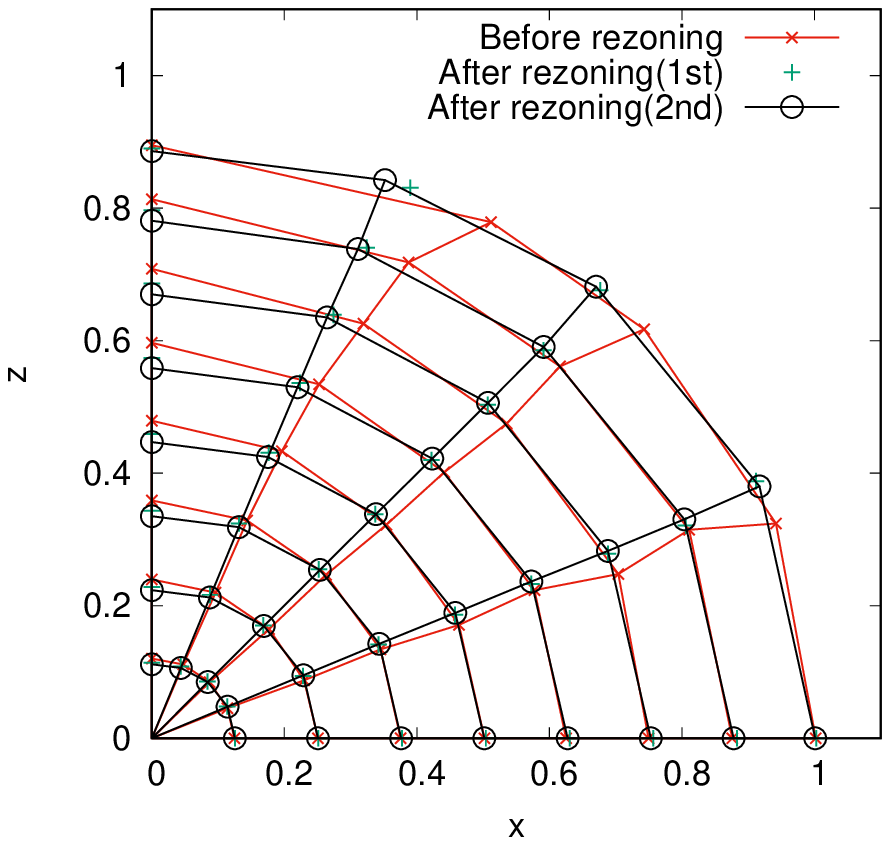} &
  \includegraphics[width=7.5cm]{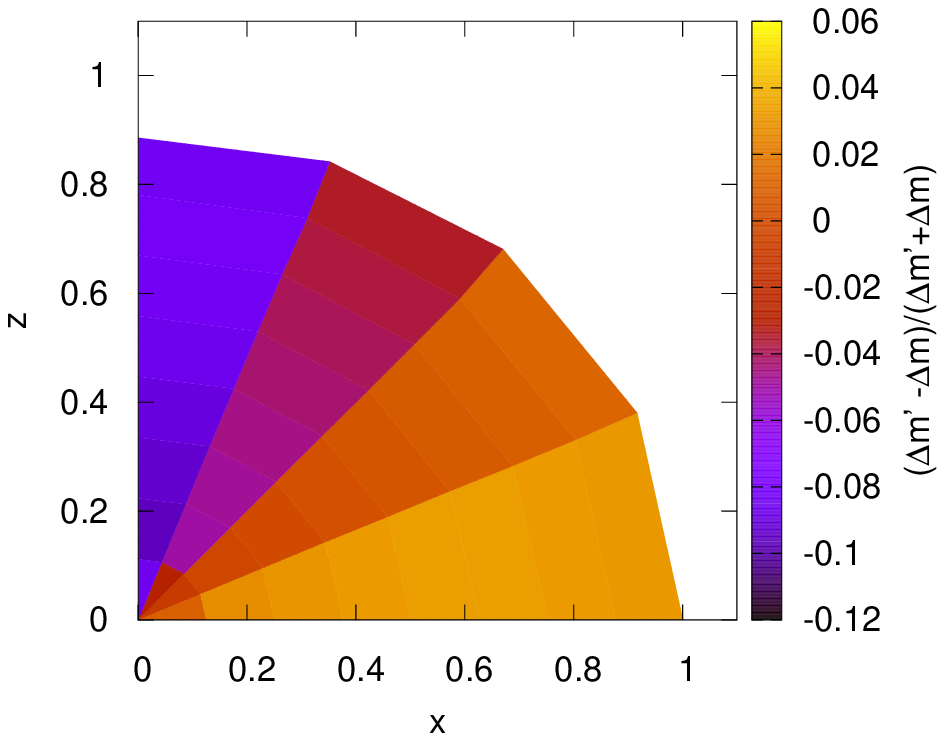}
 \end{tabular}
 \caption{(a) Configurations before and after rezoning.  One may find a jagged configuration depending on the mass distribution as the red crosses.  The rezoning finds the new mass distribution to give the equally-spaced spherical coordinates. 
 (b) The mass distributions before and after rezoning are compared in the configuration~(a).  The mass of each element increases in the red-colored region, while it decreases in the blue-colored region.
 }
 \label{fig:rezone}
\end{figure}
In the Lagrangian formulation,
we solve the coordinates~$r$ and $\theta$
with a given mass inside a finite-element fixed.
As a result, we may obtain a jagged configuration
expressed as the red crosses in Fig.~\ref{fig:rezone}~(a).
The density distribution computed through Eq.~\eqref{eq:density_def}given a certain mass distribution satisfies the Einstein and Euler equations. 
In principle, it is possible to find
the mass distribution 
that gives the equally-spaced spherical coordinates
showing the same density distribution.
Thus, we find the mass distribution as follows.
\begin{enumerate}
 \item The surface of the star at equally-spaced angle~$\theta'$ is  interpolated using the three nearby elements from Eqs.~\eqref{eq:iso2nd_x} and ~\eqref{eq:iso2nd_f}.
 \item Based on the above surface-fitted curve,
       the stellar interior is equally divided by new coordinates~$r'$ and $\theta'$.
 \item The energy density distribution
       at the coordinates~$\veps(r',\theta')$
       is computed by the isoparameteric interpolation, i.e.,
       Eqs.~\eqref{eq:xiso},
       \eqref{eq:yiso} and \eqref{eq:fiso},
       from the jagged solution.
 \item The volume~$\Delta V'$ and
       the density $\rho'$ in the new coordinates~$r'$ and $\theta'$
       are computed
       provided a new mass distribution~$\Delta m'$.
 \item By comparing the energy density~$\veps'(r',\theta')$
       directly from $\rho'$
       with the interpolated energy density~$\veps(r',\theta')$,
       a set of nonlinear equations 
       for the new mass distribution~$\Delta m'$ is obtained:
       \begin{eqnarray}
	\veps'_{jk}(\Delta m'_{11},\cdots,
	 \Delta m'_{lm},\cdots,\Delta m'_{N_{r}N_{\theta}}) = \veps_{jk},
       \end{eqnarray}
       where ($j,l$) and ($k,m$) denote the indeces for radial and polar angle directions, respectively.
 \item $\Delta m_{jk}'$ are solved by the W4 method
       with the LH decomposition,
       which slightly changes the total baryon mass
       by this process.
       To keep the total baryon mass, $\Delta m_{jk}'$ is rescaled
       as $\Delta m_{jk}' \times M /M'$ 
       where $\displaystyle M\equiv\sum_{j,k}\Delta m_{jk}$ and
       $\displaystyle M'\equiv\sum_{j,k}\Delta m_{jk}'$.
 \item The specific angular momenta
       are obtained by the interpolation from the distribution
       before rezoning.
       They are also rescaled to keep the total angular momentum
       as done in finding the mass distribution.
 \item Using the new mass distribution~$\Delta m_{jk}'$,
       the coordinates are solved to satisfy 
       the Euler and Einstein equations.
       The rezoning will be repeated
       if the obtained coordinates are deviated
       from the equally-spaced
       spherical coordinates because of the nonlinearity.
\end{enumerate}

In Fig.~\ref{fig:rezone}~(a), the red-crosses show the initially jagged configuration before rezoning.
The first rezoning gives the green-pluses and the second one
gives the black-circles.
By a few rezonings, it settles down to the equally-spaced configuration as expected.
In general, the rezoning mixes the matter inside the star
at the above procedure~(v).
Fig.~\ref{fig:rezone}~(b) displays the normalized mass difference~$(\Delta m_{jk}'-\Delta m_{jk})/(\Delta m_{jk}'+\Delta m_{jk})$ in between initial and final configurations,
which indicates that such a mass transfer indeed makes the grid equally-spaced.



\bsp	
\label{lastpage}
\end{document}